\documentstyle [psfig,10pt] {article}
\def\l {\lambda } 
\def \t {\theta }

\def\a {\alpha }

\def \d {\delta }
\def \D {\Delta }

\def \g {\gamma }
\def \G {\Gamma }

\def \b {\beta }
\def \S {\Sigma }
\def \s {\sigma }
\def \e {\epsilon }
\def \ud { {1 \over 2} }

\def \qslash {Q \kern-.5em\slash }
\def \pslash {p \kern-.5em\slash }
\def \ppslash {p' \kern-.5em\slash }
\def \Pslash {P \kern-.5em\slash }
\def \kslash {k \kern-.5em\slash }
\def \bea {\begin{equation}}
\def \eea {\end{equation}}
\def \FFL {{ {\l '}_{Jjk}^\star \l ' _{J'jk}\over (4\pi )^2 } }
\def \FFQ {{  {\l '}_{iJk}^\star \l ' _{iJ'k}\over (4\pi )^2 } }
\def \FFP {{ {\l '}_{ijJ'}^\star \l ' _{ijJ}\over (4\pi )^2 } }

\def \FFIL {{  {\l }_{iJk}^\star \l _{iJ'k}\over (4\pi )^2 } }
\def \FF2L {{ {\l }_{ijJ} \l _{ijJ'}^\star \over (4\pi )^2 } }
\def \FLU {{ {\l ' }_{iJ'k} \l  _{iJk}^{'\star } \over (4\pi )^2 } }

\def \pr  { Phys. Rev. }
\def \np { Nucl. Phys. }

\def \prl { Phys. Rev. Lett. } 
\def \pl { Phys. Lett. }

\input psfig

\if@twoside\oddsidemargin25mm
\else\oddsidemargin25mm\evensidemargin25mm\marginparwidth25mm\fi
\begin{document}
\begin{titlepage}
%\draft
%\address{Service de Physique Th\'eorique 
%CE-Saclay F-91191 Gif-sur-Yvette,  Cedex FRANCE }
\title{ Broken R parity  contributions to
 flavor changing  rates and CP   asymmetries 
 in fermion pair production at leptonic  colliders 
 \thanks 
{\it Supported by the 
 Laboratoire de la Direction des Sciences
de la Mati\`ere du Commissariat \`a l'Energie Atomique }    }
\author{{M. Chemtob and G. Moreau}\\ \\ 
{\em Service de Physique Th\'eorique}\\
{\em CE-Saclay F-91191 Gif-sur-Yvette, Cedex FRANCE}}
\maketitle
\begin{abstract}
We examine the effects of the  R parity odd renormalizable interactions 
on flavor changing rates  and CP asymmetries  in the production of 
fermion-antifermion pairs at leptonic  (electron and muon)  colliders.
In the reactions, $l^-+l^+\to f_J +\bar f_{J'}, \
[l=e , \ \mu ;  \  J\ne J' ]$ 
the  produced fermions may be  leptons, down-quarks   or up-quarks,
and the center of mass energies may range  from
the Z-boson  pole up to $ 1000$ GeV.
Off the  Z-boson pole, the flavor changing rates  are controlled by tree  level 
amplitudes and the  CP asymmetries
by interference  terms between tree and loop level  amplitudes. At the 
Z-boson pole, both observables involve loop amplitudes.
The  lepton number violating interactions,
associated with the coupling constants, $\l_{ijk} , \ \l '_{ijk}$, are only taken into account.
The consideration of  loop amplitudes  is
restricted  to the  photon and Z-boson vertex corrections.
We  briefly review flavor violation physics at colliders.
We present numerical  results  using a single, species and  family independent, 
mass parameter, $\tilde m$,  
for all the scalar superpartners and considering simple assumptions for
the family dependence of the R parity odd coupling constants. 
Finite non diagonal  rates (CP asymmetries) entail non vanishing  products of two (four) different coupling constants in different family configurations.  For lepton pair  production,   the Z-boson decays  branching ratios, $B_{JJ'}= B(Z\to l^-_J+l^+_{J'})$, scale in order of magnitude as, 
$ B_{JJ'}\approx ({\l \over 0.1})^4 ({100 GeV \over \tilde m} )^{2.5}
 \ 10^{-9} $, with coupling constants $\l = \l_{ijk} $ or 
 $\l '_{ijk}$ in appropriate family 
configurations.  The corresponding results for d- and u-quarks 
are larger, due to an extra color factor, $N_c=3$.   
The  flavor non diagonal rates, at  energies  well above the 
Z-boson pole,  slowly decrease  with  the  center of mass energy and scale  with  the 
mass parameter approximately as,
 $ \s_{JJ'}\approx  ({\l \over 0.1})^4 ({100 GeV \over \tilde m})^{2 \ - \ 3}
(1 \ - \ 10)  fbarns$.  Including the contributions from 
an sneutrino s-channel exchange 
could raise the rates  for leptons or d-quarks by one order of magnitude.  
The CP-odd asymmetries at the Z-boson pole, ${\cal A}_{JJ'}={B_{JJ'}-B_{J'J}\over B_{JJ'}+B_{J'J} } $, vary inside the range, 
 $(10^{-1}\ -\  10^{-3}) \sin \psi $, where $\psi $ is the CP-odd phase.
At energies  higher   than the Z-boson pole,  CP-odd asymmetries  for 
 leptons,  d-quarks and u-quarks pair production  lie 
approximately  at,  $(10^{-2} \ - \ 10^{-3}) \sin \psi $,
irrespective  of whether  one deals with   light or heavy flavors. 
\end{abstract}

{Saclay preprint T98/056}

{\it hep-ph/9806494}

Phys. Rev. Reference: {\bf DT6495}
\end{titlepage}
\section{Introduction}
\label{secintro}
An approximate  R parity  symmetry could greatly enhance our insight 
into the  supersymmetric flavor problem. 
As is known,  the  dimension four R parity odd 
superpotential trilinear  in the quarks and leptons superfields, 
\begin{equation}
W_{R-odd}=\sum_{i,j,k} \bigg (\ud \l _{ijk} L_iL_j E^c_k+
\l ' _{ijk} Q_i L_j D^c_k+
\ud \l '' _{ijk} U_i^cD_j^cD_k^c   \bigg )  \  , 
\label{eqi1}
\end{equation}
adds new  dimensionless  couplings
in the family  spaces of the  quarks and leptons and their superpartners.  
Comparing with  the analogous situation for the  Higgs-meson-matter
Yukawa interactions, one naturally expects  the set of 45  dimensionless 
 coupling constants, $\l_{ijk} = -\l_{jik}, \ 
\l '_{ijk} , \ \l ''_{ijk} =- \l ''_{ikj}$, to exhibit  a non-trivial  
hierarchical structure  in  the families spaces.
Our goal in this work will be  to examine a particular  class of tests 
at  high energy colliders by which one could 
access a  direct information on the family structure of  these coupling constants.

The R parity symmetry has inspired a vast literature since the 
pioneering period  of the early 80's
\cite{aulak,zwirner,hallsuz,lee,ellisvalle,rossvalle,dawson,masiero} 
and the maturation period of the   late 80's  and early 90's
\cite{dimohall,bargerg,dreiner1,ibross,hinchliffe,moha}. This subject 
 is currently witnessing a renewed  interest \cite{reviews,reviews2}. 
As is well known, the R parity odd interactions can contribute at tree level,
by exchange of the scalar superpartners, 
to processes   which violate the  baryon and lepton  numbers  as well as the  leptons and 
quarks flavors. The major part of the  existing  experimental
constraints on coupling constants is formed from the  indirect  bounds 
gathered  from the low energy phenomenology.
Most often, these have been derived on the basis 
of the  so-called  single coupling  hypothesis, 
where a single  one of the  coupling constants is assumed to 
dominate over all the others, so that
each  of the  coupling constants  contributes once at a time.
Apart from a few  isolated  cases, 
the typical bounds derived under this assumption,  
 assuming  a linear dependence  on the superpartner masses, are of order,   
$[ \l ,\  \l' ,\  \l '' ] <  ( 10^{-1} - 10^{-2}) 
  { \tilde  m\over 100 GeV }$. 

One important variant of the 
single coupling  hypothesis can be defined by assuming that the 
dominance of single operators applies at the level of the gauge (current) 
basis fields rather than the mass eigenstate fields, as was implicit in the 
above original version. 
This  appears as a more natural  assumption in models where 
the presumed hierarchies in coupling constants originate from physics at higher 
 scales (gauge, flavor, or strings).  Flavor changing  
contributions  may then be induced even when 
a single  R parity odd coupling constant is assumed to  dominate \cite{agashe}. 
While the redefined mass basis superpotential  may then  depend
on the various  unitary transformation matrices, $V_{L,R}^{u, d} $,
\cite{ellis98}, two distinguished predictive 
choices  are those where the  generation mixing  is represented
solely  in terms of the CKM (Cabibbo-Kobayashi-Maskawa)  matrix, with 
flavor changing effects  appearing in  either up-quarks 
or down-quarks flavors
\cite{agashe}. A similar situation holds for leptons with respect to the couplings, $\l _{ijk}$, and transformations, $V_{L,R}^{l,\nu } .$

A large set of constraints has also been obtained 
by  applying an extended hypothesis
of  dominance  of coupling constants by pairs (or more).
 Several analyses  dealing with  hadron flavor changing effects 
 (mixing parameters for the neutral light and
 heavy  flavored  mesons, rare mesons   decays  such as,
 $K\to \pi +\nu +\bar \nu $, ...)  \cite{agashe};
 lepton flavor changing  effects (leptons decays, $l_l^\pm \to l_k^\pm+l_n^-+l^+_p, $ \cite{roy}
 $ \quad \mu ^- \to  e^-$  conversion  processes, \cite{kim}, neutrinos Majorana mass \cite{godbole}, ...); lepton 
number violating effects (neutrinoless double beta decay 
\cite{hirsch,babu,hirsch1}); or  baryon number violating  
effects (proton decay partial branchings \cite{smirnov},
 rare non-leptonic decays of heavy mesons \cite{carlson},
nuclei desintegration \cite{carlson2},...)   have   led to 
strong bounds on a large number of quadratic  products  of the coupling constants. All of the above low energy works, however, suffer  
from one  or other form of  model dependence, whether they   rely   on the consideration of loop diagrams  \cite{smirnov},
on additional assumptions concerning the flavor mixing \cite{agashe,roy,kim},
or  on  hadronic  wave functions inputs \cite{carlson,carlson2}.

Proceeding  further with a linkage of R parity with 
physics beyond the standard model, our main  observation  in this work 
is that the R parity odd coupling constants  could by themselves 
be an independent  source of CP violation. 
Of course, the idea that the RPV interactions could act as a source of 
superweak CP violation is not a new one in the supersymmetry literature.
The principal motivation is that, whether the  RPV interactions operate by themselves or in  association with the gauge interactions, by 
exploiting  the  absence   of  strong  constraints  on 
violations with respect to the flavors of  quarks, leptons  and the
scalar superpartners by the RPV interactions, one could greatly enhance  
the potential for observability of CP violation. 
To our knowledge, one of the earliest discussion of this possibility  is contained in  ref.\cite{masiero}, where the r\^ole
 of a relative  complex phase in a pair 
of $\l'_{ijk}$ coupling constants was analyzed
in connection with  the neutral $ K, \ \bar K $ 
mesons  mixing and decays  and also with the neutron 
electric dipole moment.   This subject has attracted
increased interest in the recent literature \cite{grossman,kaplan,abel,guetta,jang,chen,adhikari,shalom,shalom2,handoko}. Thus, 
the r\^ole of complex  $\l'_{ijk} $ coupling constants 
was considered in an analysis of the muon 
polarization in  the decay, $ K^+ \to \mu^ + +\nu +\g $  \cite{chen}, 
and also of  the neutral $ B, \ \bar B$ meson CP-odd decays asymmetries \cite{kaplan,guetta,jang}; 
that of complex $\l_{ijk} $ interactions was considered 
in a study  of the spin-dependent asymmetries of 
sneutrino-antisneutrino resonant production of $\tau -$lepton pairs,
$l^-l^+\to \tilde \nu , \bar {\tilde \nu } \to \tau ^+ \tau ^-$  
\cite{shalom};  and that of  
complex  $\l''_{ijk}$   interactions was considered 
as a possible explanation for 
the cosmological baryon asymmetry \cite{adhikari},  
as well as  in the  neutral $B , \ \bar B$  decays asymmetries \cite{jang}. 
An interesting alternative proposal \cite{abel} 
is to embed the CP-odd phase in the 
scalar superpartner interactions corresponding  to interactions of 
$A'_{ijk} \l'_{ijk}$ type.
Furthermore, even if one  assumes that the 
R parity odd interactions are  CP conserving, these could  still lead, 
in combination with the  other possible  sources  of complex phases 
in the minimal supersymmetric standard model, to new tests of  CP violation.  Thus, in the hypothesis  of pair of dominant coupling constants 
new contributions involving the   coupling constants $\l '_{ijk}$ 
and the CKM complex phase can arise for  CP-odd  observables associated with  the  neutral mesons mixing parameters and decays
\cite{kaplan,guetta,jang}.  Also, 
through the interference with the  extra CP-odd phases present 
in the soft supersymmetry parameters,  $A$, 
the interactions $\l_{ijk} $ and $\l '_{ijk}$
may induce  new contributions to  electric dipole moments  \cite{hamidian}.

We propose in this work to examine  the effect that 
R parity odd CP violating  interactions could have  on  flavor non-diagonal rates
and CP asymmetries in  the production at high energy colliders 
of fermion-antifermion pairs of different 
families. We consider the two-body  reactions, $l^-(k)+l^+(k')\to f_J(p)
+\bar f_{J'}(p'), \ [J\ne J']$ where $l $ stands for electron or muon, 
the produced fermions are leptons, down-quarks   or up-quarks and
the center of mass energies span the 
relevant range of existing and planned leptonic  (electron  or muon)  colliders, namely,
from the Z-boson  pole up to $ 1000 $ GeV.  High energy colliders tests of the 
RPV contributions to the flavor diagonal reactions were
recently examined in \cite{choud96,kalin12,kalin13,kalin14} and for flavor non-diagonal reactions in \cite{mahanta}.

The physics of CP non conservation at high energy colliders  has motivated 
a  wide variety  of proposals in the past \cite{cpcoll}  and is currently
the focus of important activity.
In this work we shall limit ourselves to 
the simplest kind of observable,  namely,  the spin independent observable
involving  differences in rates  between a given flavor
 non-diagonal process and its CP  conjugated process. 
While the R parity odd interactions contribute to 
flavor changing amplitudes already at tree level, their contribution to 
spin  independent CP-odd observables entails the consideration of loop diagrams. Thus, the 
  CP asymmetries in the  Z-boson pole branching fractions,
$B (Z\to f_J+\bar f_{J'}) $, are controlled by a complex phase 
interference between non-diagonal flavor contributions to loop amplitudes,  
whereas the off Z-boson pole asymmetries  are controlled instead  by 
a complex phase 
interference  between tree and loop  amplitudes. Finite contributions 
at tree level order can arise for spin dependent CP-odd observables, as 
discussed in refs.~\cite{shalom,shalom2}.

It is useful to recall at this point that contributions
in the standard model  to the  flavor
changing rates and/or CP asymmetries   can only appear
through  loop diagrams involving the
quarks-gauge bosons interactions.  Corresponding contributions involving
squarks-gauginos or  sleptons-gauginos interactions also arise
in the minimal supersymmetric standard model. In 
studies performed some time ago
within the  standard model,  the flavor non diagonal  vector bosons (Z-boson and/or W-bosons) decay rates  asymmetries   \cite{axelrod,clements,ganapathi} and  CP-odd 
asymmetries \cite{bernabeu,hou} were found to be  exceedingly small.
(Similar conclusions were reached  in top-quark phenomenology \cite{eilam}.)
On the other hand, in most proposals of physics beyond the standard model, 
the prospects for observing flavor changing effects in 
rates \cite{axelrod,clements,ganapathi,bernabeu,hou,atwood} or in CP asymmetries \cite{cpcoll,atwood1} are on the  optimistic side. Large effects
  were  reported for the supersymmetric  corrections
in flavor changing Z-boson decay rates arising from squarks flavor mixings  \cite{duncan},   but the conclusions from this initial work have been 
challenged in a subsequent work \cite{mukho} involving a more complete calculation.

The possibility that the R parity odd interactions could  contribute 
to the CP asymmetries at observable  levels 
depends  in the first place on the accompanying mechanisms 
responsible for the flavor changing rates. Our working assumption in this work will  be
that the R parity odd interactions are the dominant contributors 
to flavor non-diagonal amplitudes.

The contents of this paper  are organized into 
4 sections. In Section \ref{sec1}, 
we develop the basic formalism for describing 
the scattering  amplitudes 
at tree and one-loop levels. We discuss the case of leptons, 
down-quarks and up-quarks 
successively in subsections \ref{subsec1}, \ref{subsec2} and  \ref{subsec3}. 
The evaluation of  the one-loop  loop 
diagrams is based on the standard formalism 
of \cite{pasvel}.  Our calculations here  closely parallel
similar ones developed \cite{ellis,bhatta3} in  connection with corrections  to the 
Z-boson partial widths.  In Section  \ref{sec5}, we first  
briefly review the physics of flavor violation  and next present our numerical results
for the  integrated cross sections (rates)  and   CP asymmetries for fermion pair 
production at and off the Z-boson pole. In Section \ref{seconc}, we state our main conclusions 
and  discuss the impact of 
our results on possible experimental measurements.  

\section{Production of fermion pairs of different flavors}
\begin{figure} [h]
\begin{center}
\leavevmode
\psfig{figure=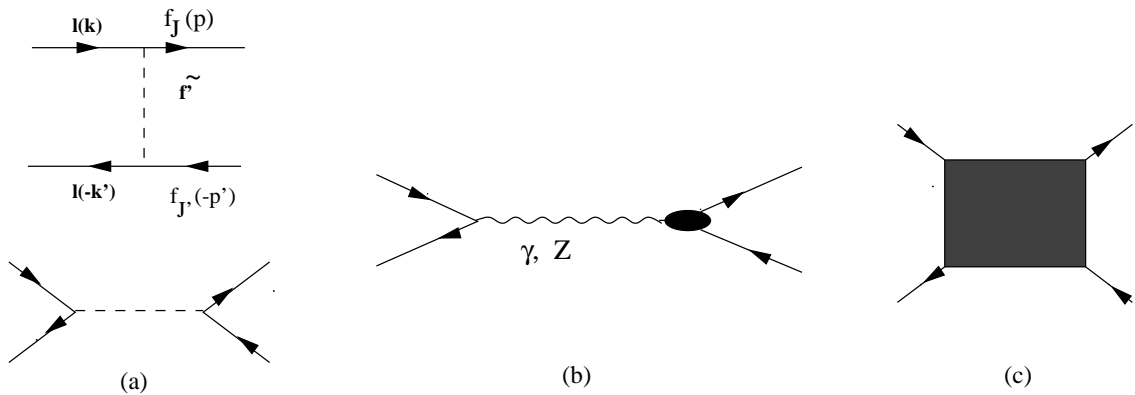}
%\epsfbox{fig1.eps}
\end{center}
%\vskip 0.5cm
\caption{Flavor non-diagonal process of $l^-l^+$ production of 
a fermion-antifermion pair, $l^-(k)+l^+(k')\to f_J (p) +\bar f_{J'} 
(p') $. The tree level diagrams in $(a)$ represent t- and s-channel exchange 
amplitudes. 
The  loop   level diagrams represent
$\g $  and $ Z $  gauge boson exchange amplitudes with dressed vertices 
in $(b)$  and box amplitudes in $(c)$.}
\label{fig1}
\end{figure}

\vskip 0.5cm
\label{sec1}
In this section  we shall examine the contributions induced by 
the RPV (R parity violating)  couplings on the flavor changing processes, 
$l^-(k)+l^+(k')\to f_J(p)
+\bar f_{J'}(p'), \ [l=e,\ \mu ; \ J\ne J']$ where $f$ stands for leptons or quarks and 
$J,  \   J' $ are family  indices.  The relevant 
tree and one-loop level diagrams are shown schematically in Fig. \ref{fig1}.
%%%%%%%%%%%%%%%%%%%%%%%%%%%%%%%%%%%%%%%%%%%%%%%%%
%%%%%%%%%%%%%%%%%%%%%%%%%%%%%%%%%%%%%%%%%%%%%%%%%
At one-loop order, there arise $\g  -$ and  $Z-$
boson exchange triangle diagrams as well as  box  diagrams.  
In the sequel, for clarity,  we shall 
present the formalism for the one-loop contributions
only for the dressed $Z f\bar f$ vertex in the
Z-boson exchange amplitude.  The dressed  $\g $-exchange amplitude has 
a similar structure and  will be added in together 
with the Z-boson exchange  amplitude at the level of the  numerical results.
Since we shall repeatedly refer in the text to the 
R parity odd effective Lagrangian  
for the fermions-sfermion Yukawa interactions, we quote below its full expression,
\begin{eqnarray}
L&=&\sum_{ijk} \bigg \{ \ud \l_{ijk}[\tilde  \nu_{iL}\bar e_{kR}e_{jL} +
\tilde e_{jL}\bar e_{kR}\nu_{iL} + \tilde e^\star _{kR}\bar \nu^c_{iR} e_{jL}
-(i\to j) ]  \cr
&+&\l '_{ijk}[\tilde  \nu_{iL}\bar d_{kR}d_{jL} +
\tilde d_{jL}\bar d_{kR}\nu_{iL} + \tilde d^\star _{kR}\bar \nu^c_{iR} d_{jL}
-\tilde  e_{iL}\bar d_{kR}u_{jL} -
\tilde u_{jL}\bar d_{kR}e_{iL} - \tilde d^\star _{kR}\bar e^c_{iR} u_{jL}
]  \cr 
&+&\ud {\l ''}_{ijk}\e_{\a \b \g }[\tilde  u^\star _{i\a R}\bar d_{j\b R}d^c_{k\g L} +
\tilde  d^\star _{j\b R}\bar u_{i\a R}d^c_{k\g L} +
\tilde  d^\star _{k\g R}\bar u_{i\a R}d^c_{j\b L}  -(j\to k)] \bigg \} +h.c. \ ,
\label{eqyuk}
\end{eqnarray}
noting that the summations run  over the (quarks and leptons) 
families indices, 
$i,j,k  =[(e,\mu , \tau ); \  (d,s,b); \ (u,c,t)]$, subject to the antisymmetry properties, $\l_{ijk}=-\l_{jik}, \   \l ''_{ijk}=-\l '' _{ikj}$.
We use precedence conventions for operations on Dirac spinors such that
charge conjugation acts first, chirality projection second and Dirac bar 
third, so that, $\bar \psi ^c_{L,R}= \overline {(\psi ^c)_{L,R} }.$
\subsection{ Charged lepton-antilepton pairs}
\label{subsec1}
\subsubsection{General formalism}
The process  $l^-(k)+l^+(k')\to e^-_J(p) +e^+_{J'}(p')$, for 
$l=e,\  \mu ; \ J\ne J'$, can pick up a finite contribution at tree level 
from  the R parity odd couplings, 
$\l_{ijk}$, only. For clarity, we treat in the following the case 
of electron colliders, noting that the case of muon colliders is 
easily deduced by replacing all occurrences in the RPV coupling constants 
of the index $1$ by the index $2$.
There occur both  t-channel and s-channel  $\tilde \nu_{iL}$ 
exchange contributions, of the type shown by the Feynman diagrams in $(a)$ 
of  Fig. \ref{fig1}. 
The scattering amplitude at tree level, $M_t$,  reads: 
\begin{eqnarray}
&M^{JJ'}_t&= -{1\over 2(t-m^2_{\tilde \nu_{iL} } )  } 
\bigg [ \l_{i1J} \l^\star_{i1J'}
\bar u_R(p) \g_\mu v_R(p')\bar v_L(k')\g_\mu u_L(k)  \cr
&+&\l ^\star _{iJ1} \l _{iJ'1}
\bar u_L(p) \g_\mu v_L(p')\bar v_R(k')\g_\mu u_R(k) \bigg ] \cr
&-& {1 \over s-m^2_{\tilde \nu_{iL} } }\bigg [\l _{i11}\l ^\star _{iJJ'}
\bar v_R(k') u_L(k) \bar u_L(p) v_R(p') +
\l _{i11}^\star \l _{iJJ'}
\bar v_L(k') u_R(k) \bar u_R(p) v_L(p') \bigg ] ,
\label{eq1}
\end{eqnarray}
where to obtain the saturation structure in the 
Dirac spinors indices for the t-channel terms, 
we have applied the  Fierz rearrangement formula, 
\hfill  $\bar u_R(p) u_L(k) \bar v_L(k')v_R(p') =\ud 
\bar u_R(p) \g_\mu v_R(p')\bar v_L(k')\g_\mu u_L(k) .$
The   t-channel (s-channel) exchange terms
on the right hand side of eq.(\ref{eq1}) include two terms each, 
called ${\cal R}$- and ${\cal L}$-type, respectively. These two terms   differ 
by a chirality flip,  $L\leftrightarrow  R $,  and correspond 
to the distinct diagrams  where the  exchanged
sneutrino is emitted or absorbed at the upper (right-handed) vertex.

The Z-boson exchange amplitude (diagram  $(b)$ in Fig. \ref{fig1})
 at loop level, $M_l$,  reads:
\begin{eqnarray}
M^{JJ'}_l = \bigg ({g\over 2 \cos \t_W}\bigg )^2 \bar v (k')\g_\mu \bigg  (a(e_L) P_L+a(e_R)P_R \bigg )u(k) 
{1\over s-m_Z^2+im_Z\G_Z} \G_\mu ^Z (p,p') , 
\label{eq2}
\end{eqnarray}
where the Z-boson current amplitude   vertex function, $\G^Z_\mu  (p,p')$, 
is defined through  the effective Lagrangian density, 
$$ L= -{g\over 2 \cos \t_W } Z^\mu  \G_\mu ^Z (p,p').$$ 
For later convenience, we  record for the  processes, 
$Z(P=p+p') \to f(p)+\bar f'(p')$ and $Z(P)\to \tilde f_H(p)  +
\tilde f_H ^\star  (p')$, 
the familiar definitions of the Z-boson bare vertex functions, 
\begin{eqnarray}
\G _\mu ^Z(p,p') =  \bigg [\bar f(p)\g_\mu \bigg  (
a(f_L)P_L +a(f_R) P_R\bigg )
f'(p')  + (p-p')_\mu  
\tilde f_H^\star (p') 
a(\tilde  f_H)\tilde f_H(p) \bigg ], 
\label{eqm1}
\end{eqnarray}
where  the quantities denoted,   $a(f_H)\equiv a_H(f)$  and 
$a(\tilde f_H) $, taking equal values for both fermions and sfermions, 
are defined by,  $ a(f_H)=a(\tilde f_H)=2T_3^H (f)-2Qx_W, $ where  
$ H=(L,R), \ x_W= \sin^2 \t_W , \ T_3^H $ are $ SU(2)_H $  Cartan subalgebra 
generators, and 
$Q= T_3^L +Y , \ Y  $ are  electric charge and weak hypercharge.  These 
parameters satisfy  the useful  relations:  
$a(\tilde f_H^\star ) =-a(\tilde f_H), \  a_L(f^c)=-a_R(f), 
\  a_R(f^c)=-a_L(f).$ 
Throughout this paper we shall use the conventions in Haber-Kane review 
\cite{haber} (metric signature  $(+---)$, 
$P_{L\choose R}=(1\mp \g_5)/2,   $ etc...) and adopt 
the  familiar summation convention on dummy indices. 

The  Lorentz covariant  structure of the dressed Z-boson  current amplitude
in the process,
$Z(P)\to f_J(p) +\bar f_{J'}(p')$,  for a generic value of 
the Z-boson invariant mass $s=P^2$, involves 
three pairs  of vectorial and tensorial vertex functions, 
which are defined in terms of  the general decomposition:
\begin{eqnarray}
 \G _\mu ^Z(p,p')&=&
\bar u(p) \bigg [\g_\mu 
\bigg (\tilde A_L^{JJ'} (f)P_L+
\tilde A_R^{JJ'}(f) P_R\bigg ) \cr 
&+& {1\over m_J+m_{J'}} 
\s_{\mu \nu } \bigg ( (p+p')^\nu  [ia^{JJ'}  + \g_5 d^{JJ'} ] 
+(p-p')^\nu  [ ib^{JJ'} +\g_5 e^{JJ'} ] 
\bigg ) \bigg ]  v(p')  \ , \cr
&&
\label{eq3}
\end{eqnarray}
where,  $ \s_{\mu \nu } = {i\over 2} [\g_\mu , \g_\nu ] $. 
The vector vertex functions  separate additively into
the classical (bare)  and loop contributions, 
$\tilde A_H^{JJ'}(f)=a_H(f)\d_{JJ'}  +A_H^{JJ'}(f), \ [H=L,R]$. 
The  tensor vertex functions, associated with $\s^{\mu \nu } (p+p')_\nu $,  include the familiar magnetic and 
electric $Z \ f\  \bar f$ couplings, such that  the flavor diagonal vertex functions,
$-{g\over 2 \cos \t_W} {1\over 2m_J} [a^{JJ}, d^{JJ} ]$,   
 identify, in the small momentum transfer limit, 
with  the fermions Z-boson current 
magnetic and  (P and CP-odd) electric dipole moments, respectively. 
In working with  the spinors matrix elements,
 it is helpful to recall the mass shell relations,
\hfill    $ \bar u (p)  \pslash = m_J \bar u(p) , \ \quad  
\pslash '  v(p') =- m_{J'}   v(p')$, and
the Gordon type identities, appropriate to the saturation of 
the Dirac spinor indices, $\bar u(p) \cdots v(p')$,  
$$ \bigg [(p\pm p')_\mu  {\g_5 \choose 1} 
+i\s_{\mu \nu }
(p\mp p')^\nu   {\g_5 \choose 1} \bigg ]= (m_J+m_{J'}) \g_\mu {\g_5 
\choose 1} ,$$
$$ \bigg [(p\mp p')_\mu  {\g_5 \choose 1} 
+i\s_{\mu \nu }
(p\pm p')^\nu   {\g_5 \choose 1} \bigg ]= (m_J-m_{J'}) \g_\mu {\g_5 
\choose 1 } . $$
Based on these identities, one also checks that the additional 
vertex functions,  $[b^{JJ'} , \ e^{JJ'}]$,
associated with the Lorentz covariants,  $\s^{\mu \nu } (p-p')_\nu  [1, \g_5]$, 
can be expressed as linear combinations of the 
vector  or  axial covariants, $\g_\mu \  [1,\g_5] $,  and the total 
momentum covariants, $(p+p')_\mu \  [1,\g_5] $. The latter will yield, upon contraction 
with the initial state $Zl^-l^+$ vertex function,
to  negligible mass terms in the initial leptons.

Let us now perform the summation 
over the  initial and final states  polarizations for the summed tree and 
loop amplitudes, $M^{JJ'}= M_t^{JJ'}+M_l^{JJ'}$,
 where the lower suffices $t , l $ stand 
for tree and loop, respectively. 
(We shall not be interested in this work  in spin observables.)
A straightforward calculation, carried out for the
squared sum of the tree and loop amplitudes, yields the result 
(a useful textbook to consult  here is ref. \cite{peskder}): 
\begin{eqnarray}
\sum_{pol}&& \vert M^{JJ'}_t+M^{JJ'}_l\vert^2= \bigg 
\vert -{  \l_{i1J} \l^\star _{i1J'} \over 
2(t-m^2_{\tilde \nu_{iL} } )}  +\bigg ({g\over 2\cos \t_W}\bigg )^2
{  a(e_L) A_R^{JJ'}(e, s+i\e ) \over 
s-m_Z^2+im_Z\G_Z   } \bigg \vert^2 16 (k\cdot p)(k'\cdot p') \cr &+&8m_Jm_{J'} 
(k\cdot k') \varphi_{LL} ({\cal R})   +8m_e^2 (p\cdot p') 
\varphi_{RR}({\cal R}) + \cr 
&+& \bigg \vert -{  \l_{iJ1} \l^\star _{iJ'1} \over 
2(t-m^2_{\tilde \nu_{iL} } )}  +\bigg ({g\over 2\cos \t_W}\bigg )^2
{  a(e_R) A_L^{JJ'}(e, s+i\e ) \over 
s-m_Z^2+im_Z\G_Z   } \bigg \vert^2 16 (k\cdot p)(k'\cdot p')\cr 
&+&8m_Jm_{J'} 
(k\cdot k') \varphi_{RR}({\cal L})   +
8m^2_e 
(p\cdot p') \varphi_{LL}({\cal L})  
+8\bigg  \vert {\l_{i11} \l ^\star _{iJJ'} \over 
s-m^2_{\tilde \nu_{iL} } } \bigg \vert ^2 (k\cdot k') (p\cdot p'),  
\label{eqr4}
\end{eqnarray}
where we  have introduced the following functions,
associated with the ${\cal R} $- and ${\cal L}$-type contributions:
\begin{eqnarray}
\varphi_{HH'} ({\cal R}) &=& -\bigg ({g\over 2\cos \t_W}\bigg )^2
\bigg ({ a(e_H) A_{H'}^{JJ'}(e, s+i\e ) \over 
s-m_Z^2+im_Z\G_Z }\bigg )^\star  \bigg ({ \l_{i1J}\l_{i1J'}^\star \over 
2(t-m^2_{\tilde \nu_{iL} } )  } \bigg ) +c.\ c\  \ , \cr  
\varphi_{HH'} ({\cal L}) &=& -\bigg ({g\over 2\cos \t_W}\bigg )^2
\bigg ({ a(e_H) A_{H'}^{JJ'}(e, s+i\e ) \over 
s-m_Z^2+im_Z\G_Z }\bigg )^\star  \bigg ({ \l_{iJ1}\l_{iJ'1}^\star \over 
2(t-m^2_{\tilde \nu_{iL} } )  } \bigg ) +c.\ c\  \ . 
\label{eq4p}
\end{eqnarray}
The two sets of terms in
 eqs.(\ref{eq4p}) and (\ref{eqr4}), labelled by the letters, $ {\cal R},  {\cal L}$, are
 associated with the two t-channel exchange contributions 
in the tree amplitude, eq.(\ref{eq1}),  which  differ
by the  spinors chirality structure  and the substitutions, 
$\l _{i1J} \l ^\star _{i1J'} \to  
\l^\star _{iJ1} \l _{iJ'1} .$
The terminology,  ${\cal L, \ R } $, is motivated  by the fact that these
contributions are controlled by the Z-boson left and right 
chirality vertex functions, $A_L $ and $A_R$, respectively, in the massless limit.

The  imaginary shift  in the 
argument,  $s+ i \e $ (representing the upper lip of 
the cut  real axis in the complex $s$-plane) of the vertex functions,
$A_H^{JJ'}(f,s+i\e )$, has been appended 
to remind us that the
one-loop  vertex functions are complex functions 
in the complex plane of  the virtual Z-boson mass squared, 
$s=(p+p')^2$, with  branch cuts  starting at the physical thresholds 
where  the production processes, such as, $ Z\to f+\bar f $ or $Z\to \tilde  f
+\tilde f^\star $,  are raised on-shell.
For notational simplicity, we have omitted writing
several terms proportional to the initial leptons masses 
and also some of the small  subleading 
terms  arising from  the loop amplitude squared.  
At the energies of interest,  whose scale is set 
by the initial center of mass energy or by the  Z-boson mass,
 the  terms involving factors of the initial leptons masses $m_e$, 
are entirely negligible, of course.  Thus, 
the  contributions associated with $\varphi_{RR} ({\cal R}),\ 
\varphi_{LL}({\cal L})   $ 
can  safely be dropped.
Also, the contribution from $\varphi_{LL} ({\cal R}) $ and 
$\varphi_{RR} ({\cal L}) $
which are proportional to
the final state leptons masses, $m_J, $ and $  m_{J'}$, 
can to a good approximation be  neglected for leptons production.
Always in the same approximation, we find also that interference terms are absent between the s-channel 
exchange and     the 
t-channel amplitudes and between  the s-channel  tree and Z-boson  exchange 
loop    amplitudes.  Similarly,  because of the opposite 
chirality structure of the first two terms in  $M_t^{JJ'}$, 
their cross-product contributions give negligibly small mass terms.
\subsubsection{CP asymmetries}
Our main concern in this work bears on  the comparison of 
the pair of CP conjugate reactions,
$l^-(k)+l^+(k')\to e^-_J(p) +e^+_{J'}(p')$ and 
$l^-(k)+l^+(k')\to e^-_{J'}(p) +e^+_{J}(p')$. 
Denoting the summed tree and one-loop  probability  amplitudes for these reactions as,
$ M^{JJ'}= M_t^{JJ'}+M_l^{JJ'} , \ 
\bar  M^{JJ'}= M_t^{J'J}+M_l^{J'J} = M^{J'J}$, we observe that these amplitudes are simply
related  to one another  by means of a specific  complex conjugation operation. 
The general structure  of this relationship can be expressed schematically as:
\begin{eqnarray}
M^{JJ'}=a_0^{JJ'}+\sum_\a a_\a ^{JJ'}  F^{JJ'}_\a (s+i\e ), \ \ 
\bar M^{JJ'}=a^{JJ' \star } _0+\sum_\a a^{JJ'\star } _\a  
F^{J'J} _\a (s+i\e ),
\label{eqst1}
\end{eqnarray}
where for each of the  equations above,  referring to amplitudes 
for pairs of CP conjugate processes, 
the first and second terms  correspond to  the tree
and loop level contributions, with $a^{JJ'}_0 , \ a_0^{J'J}= a_0^{JJ'\star }$, 
representing   the tree  amplitudes and
$a_\a ^{JJ'},  \ a_\a ^{J'J} = a_\a ^{JJ' \star }$ 
and $F_\a ^{JJ'} , \ F_\a ^{J'J} =F_\a ^{JJ'} $ 
representing the  complex valued  coupling constants products and 
momentum integrals in the loop amplitudes.  The functions 
$F^{JJ'} $ must  be symmetric under the interchange, $J\leftrightarrow J'$.
The summation index $\a  $ labels  the  family
configurations for the intermediate  fermions-sfermions which can
run inside the  loops. 
Defining  the CP asymmetries  by the normalized  differences,
$${\cal A}_{JJ'}={ \vert M^{JJ'} \vert ^2- \vert \bar  M^{JJ'} \vert ^2 \over 
\vert M^{JJ'} \vert ^2+ \vert \bar  M^{JJ'} \vert ^2},$$ 
and inserting the decompositions in eq.(\ref{eqst1}), the result
separates additively into two types of terms:
\begin{eqnarray}
{\cal A}_{JJ'}&=&{2\over \vert a_0\vert ^2}\bigg [ 
\sum_\a  Im(a_0a_\a ^\star ) Im (F_\a (s+i\e )) \cr 
&-& \sum_{\a < \a ' } Im(a_\a a_{\a '}^\star ) Im (F_\a (s+i\e )
F_{\a '}^\star (s+i\e ) )\bigg ],
\label{eq5}
\end{eqnarray}
where, for  notational simplicity, we have suppressed the fixed external  family
indices on  $ a_{0 }^{JJ'} , \ a_{\a } ^{JJ'} $ and $ F_\a ^{JJ'}$,
and   replaced the
full denominator by the tree level amplitude, since this is expected to dominate over the loop amplitude.
The first term  in (\ref{eq5}) is associated with an interference between 
tree and loop amplitudes  and the second  with an
interference between  terms arising from different family contributions 
in the loop amplitude.
In the second term of eq.(\ref{eq5}),
the two imaginary parts  factors are antisymmetric  under  the 
interchange of indices,  $\a $ and $\a '$,  so that their  product is 
symmetric  and allows one to  
write, $\sum_{\a < \a ' } = \ud \sum_{\a \ne \a '} .$
To obtain a more explicit formula, let us  specialize 
to the specific case where the Z-boson  vertex functions  decompose as, 
$ A_H^{JJ'}(f,s+i\e ) = \sum_\a  b_{JJ'}^{H\a }
I_{H\a }^{JJ'}  (s+i\e )$.  The  first factors,   $b^{H\a }_{JJ'} = 
\l_{ijJ} \l^\star _{ijJ'}  $ (using $\a =(ij) $ and 
notations for the one-loop contributions to be 
described in the next subsection), include the CP-odd phase from 
the  R parity odd coupling constants. The  second factors,
$I_{H \a } ^{JJ'} $,  include the CP-even phase from the  unitarity cuts 
associated to the physical on-shell  intermediate states.
In the notations  of eq.(\ref{eqst1}), 
$$  a_{\a }^{JJ'}= ({g\over 2\cos \t_W })^2   a(e_{H'})  b^{H\a } _{JJ'} , \  
F_{\a }^{JJ'} =  I_{H\a } ^{JJ'}  (s+i\e )/(s-m_Z^2+im_Z\G_Z),$$
where the right hand sides
incorporate appropriate sums over the chirality indices, $H', \ H$ of the initial and final 
fermions, respectively.

Applying eq.(\ref{eqst1}) to the 
  asymmetry  integrated  with respect to the scattering angle,
one derives for the corresponding integrated  tree-loop interference contribution,
\begin{eqnarray}
{\cal A}_{JJ'}&=&-4\bigg ({ g\over 2\cos \t_W }\bigg )^2
a(e_L) Im(\l_{i1J}\l^\star _{i1J'} a_{JJ'}^{\a \star }(f_R) )
Im\bigg ({  I^R_\a (s+i\e )\over s-m_Z^2+im_Z \G_Z }\bigg  ) \cr 
&\times &\int_{-1}^{1}dx\  {(1- x)^2\over 
(2(t-m_{\tilde \nu_{iL}}^2 ) } 
\bigg [ \sum_i \vert \l_{i1J}\l_{i1J'}^\star \vert^2 
\int_{-1}^1 dx \  { (1- x)^2\over  4(t-m_{\tilde \nu^2_{iL} })^2 } 
\bigg ] ^{-1}\ , 
\label{eq6}
\end{eqnarray}
where,   $ \theta , \ [x=\cos \theta ]$ denotes  the scattering angle variable
in  the center of mass frame
and the  Mandelstam variables in the  case of 
massless final state fermions take the simplified expressions, 
$ s \equiv (k+k')^2 , \ t  \equiv (k-p)^2=  -\ud s (1- x), 
\ u \equiv  (k-p')^2   = -\ud s (1+ x)$.  
Useful kinematical relations in  the  general  case with 
final fermions masses, $m_J, \ m_{J'}$,  are: 
$\sqrt s=2 k= E_p+E_{p'}, \  t=  m_J^2 -s E_p (1-\b x) , \ u= m_{J'}^2 -s E_{p'}(1 +\b ' x), $
where, $E_p =(s+m_J^2-m_{J'}^2)/(2\sqrt s ),\ 
E_{p'}  =(s+m_{J'}^2-m_{J}^2)/(2\sqrt s ), \ \b = p/E_p,\ \b ' =p/E_{p'}$,
with  $k, \ p $  denoting  the 
center of mass  momenta of the two-body initial and final states, respectively.
The unpolarized differential cross section reads then, 
$ d\s /dx = {\vert p \vert \over 
128 \pi s \vert k \vert } \sum_{pol} \vert M \vert^2.$ 

For the Z-boson pole  observables, 
the flavor non-diagonal  branching ratios and
CP asymmetries (where one sets, $s=m_Z^2$)  are
defined   in terms of the  notations specified in the preceeding paragraph by the equations, 
\begin{eqnarray}
B_{JJ'}&\equiv & {\G (Z\to f_J+\bar f_{J'} )+ \G (Z\to f_{J'} +\bar f_J ) \over \G (Z \to all)  }
=2  {  \vert A_L^{JJ'} (f) \vert^2 +  \vert A_R^{JJ'} (f) \vert^2 
\over \sum_f \vert a_L (f) \vert^2 +  \vert a_R (f) \vert^2 } ,\cr
{\cal A}_{JJ'}&\equiv & {\G (Z\to f_J+\bar f_{J'} )- \G (Z\to f_{J'} +\bar f_J ) \over
\G (Z\to f_J+\bar f_{J'} )+ \G (Z\to f_{J'} +\bar f_J )}  \cr   &=&
- 2 {\sum_{H=L,R}\sum_{\a < \a ' } Im(b^{H\a  } _{JJ'} b^{ H\a '\star }_{JJ'} )
Im (I_{H\a }  ^{JJ'} (s+i\e ) I_{H \a '}^{JJ'\star }  (s+i\e ) ) \over 
\sum_{H=L,R}\vert \sum_\a  b_{JJ'}^{H\a } (f)
F_H^\a (s+i\e ) \vert ^2  }  \ .
\label{eq6p}
\end{eqnarray}
For completeness, we  recall the formula for the Z-boson decay width in 
fermion pairs (massless limit), 
$$\G (Z\to f_J+\bar f_{J'}) = {G_F m_Z^3 c_f\over 12 \sqrt {2} \pi }
(\vert A_L^{JJ'}(f) \vert ^2 +\vert A_R^{JJ'}(f) \vert ^2 ),$$
where, $c_f=[1,\ N_c], $ for $[f=l,q] \ (N_c=3  $ 
is the number of colors in the $ SU(3)_c $ color  group) and the experimental value  for the total width,
$\G (Z \to  all )_{exp}= 2.497 $ GeV.

The expressions in eqs.(\ref{eq6}) and (\ref{eq6p})  for the CP asymmetries
explicitly incorporate the property of these observables of  depending on 
combinations of the RPV coupling constants, such as, 
$Arg ( \l _{i1J} \l ^\star _{i1J'} 
\l^\star _{i'jJ} \l _{i'jJ'} ), $ or $Arg ( \l _{ijJ} \l ^\star _{ijJ'} 
\l^\star _{i'j'J} \l _{i'j'J'} )$,  which are invariant 
under complex phase redefinitions of the fields.  This freedom
under rephasings of the quarks and leptons superfields actually removes 
$21$ complex phases from the complete general set of $ 45$  complex 
RPV coupling constants.

\subsubsection{One-loop  amplitudes}
\begin{figure} [h]
\begin{center}
\leavevmode
\psfig{figure=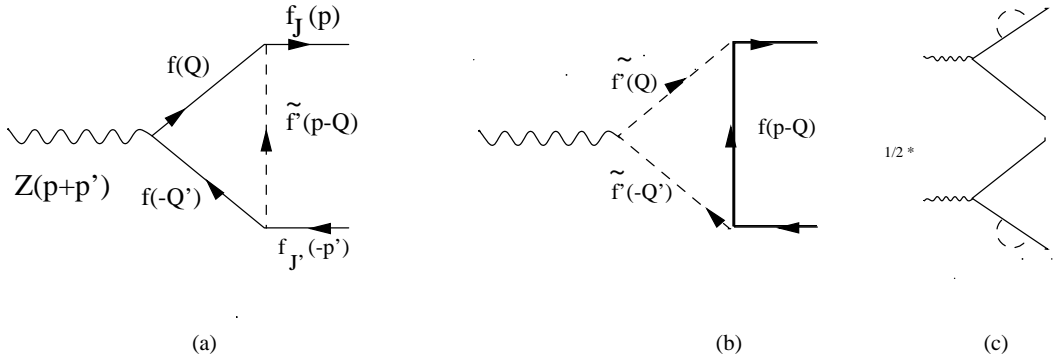}
%\epsfbox{fig2.eps}
\end{center}
\caption{One-loop diagrams for the dressed $Z(P) \ f(p) \bar f(p') $ vertex.
The flow of four-momenta for the intermediate  fermions  in $(a)$ 
is  denoted  as, 
$ Z(P)\to f(Q)+\bar f(Q') \to f_J(p) +\bar 
f_{J'} (p') $.  Similar notations are used for the sfermions diagram 
in $(b)$ where,  $ Z(P)\to \tilde f'(Q) +\tilde f ^{'\star } (Q') $, 
and for the self-energy diagrams in $(c)$.  }
\label{fig2}
\end{figure}

The relevant triangle Feynman diagrams,
which contribute to the dressed  Z-boson  leptonic vertex, 
$Z(P) l^- (p)  l^+ (p')$, appear in three types,  fermionic, scalar and self-energy, as shown in Fig.  \ref{fig2}. 
We consider first the contributions induced by the R parity odd
couplings, $\l '_{ijk} $.
%%%%%%%%%%%%%%%%%%%%%%%%%%%%%%%%%%%%%%%%%%%%%%%%%
%%%%%%%%%%%%%%%%%%%%%%%%%%%%%%%%%%%%%%%%%%%%%%%%%
The intermediate lines  can assume two distinct
configurations which  contribute both,
in the limit of vanishing external fermions masses, 
to the left-chirality vertex functions only.  We shall refer to such contributions by the name  ${\cal L}$- type contributions, 
reserving the name ${\cal R}$-type to contributions to the
right-chirality vector couplings. 
The two  allowed configurations  for the internal fermions and 
sfermions are:
$ f= {d_k \choose u_j^c} ; \tilde f'= 
{\tilde u_{jL}^\star  \choose  \tilde d_{kR}} .$
Our calculations of the triangle diagrams employ the kinematical conventions  for the flow of electric charge  and momenta 
indicated in Fig.\ref{fig2}, where $ P=p+p'=Q+Q'=k+k'$.
The  summed fermion and scalar  Z-boson current contributions are given by: 
\begin{eqnarray}
\G_\mu ^Z ({\cal L}) &=&+i  N_c {\l '}^\star _{Jjk}{\l '}_{J'jk} 
\bigg [ \int_Q 
{  \bar u(p)[P_R(\qslash +m_f )\g_\mu (a(f_L)P_L+a(f_R)P_R) (-\qslash '+m_f) P_L]v(p')
\over (Q^2-m_f^2)((Q-p-p')^2-m_f^2)((Q-p)^2-m_{\tilde f'}^2) }\cr
&+&  \int_Q 
{ a(\tilde f'_L)(Q-Q')_\mu  \bar u(p)[P_R(\pslash -\qslash +m_f )P_L] v(p')
\over (Q^2-m_{\tilde {f}'}^2)((Q-p-p')^2-m_{\tilde {f}' }^2)
((Q-p)^2-m_f^2) } \bigg ].
\label{eq7}
\end{eqnarray}
The integration measure is defined as, $ \int_Q = {1\over (2\pi )^4} \int d^4 Q$.
For a convenient derivation   of the  self-energy diagrams, one may invoke 
the on shell  renormalization condition which relates these to the
fields renormalization constants.
Defining schematically   the self-energy  vertex functions for a Dirac fermion field $\psi $
by the  Lagrangian density, $ L=i \bar \psi 
(\pslash  -m +\S (p) )  \psi , \ \S (p)=
m \s_0 +\pslash  (\s^{L} P_L+\s^{R}P_R )$,
the transition from bare  to renormalized fields and mass terms  may be 
effected by the replacements, 
$$ \psi_H \to {\psi_H\over (1+\s_{H})^\ud } = \psi_HZ_H^\ud , \ 
\  m \to  m {(1+\s^{L})^\ud (1+\s^R)^\ud \over (1-\s_0)   } .\ \  [H=L,R]$$ 
By a straightforward generalization to the case of several fields, labelled by a family index $J$,
the fields renormalization constants become matrices, 
$Z^H_{JJ'}= (1+\s^H)_{JJ'}^{-1} $. The 
self-energy contributions  to 
the dressed Z-boson vertex function is then described as,
\begin{eqnarray}
\G ^Z _{\mu }(p,p')_{SE} &=& \sum_{H=L,R}\bigg ( (Z^H_{JJ'}
Z^{H\star }_{J'J} )^\ud -1 \bigg ) \bar u(p) \g_\mu a(f_H) P_H v(p')
\cr  &=&- \sum_{H=L,R} \ud (\s^H_{JJ'} (p)+
\s^{H\star }_{J'J}(p') )  \bar u(p) \g_\mu a(f_H) P_H v(p'),
\label{eqp8}
\end{eqnarray}
where for the case at hand,
\begin{eqnarray}
\S_{JJ'}(p)&=& -i N_c {\l '}^\star _{Jjk}{\l '}_{J'jk} 
\int_Q { P_R(\qslash +m_f)P_L\over (-Q^2+m_f^2)
(-(Q-p)^2+m_{\tilde {f}'}^2 ) } ,
\label{eq8}
\end{eqnarray}
so that $\s_{JJ'}^R=0$ and $\s_0=0$. Similar
Feynman graphs to those of  Fig. \ref{fig2}, 
and similar formulas to those of eqs.(\ref{eq7}) and (\ref{eqp8}),
obtain for the dressed photon  current case, $\g (P) l^-(p)  l^+(p')$.

We organize our one-loop calculations 
in line with the  approach  developed by 
't Hooft and Veltman \cite{hooft}  and Passarino and  Veltman\cite{pasvel},
keeping in  mind that our spacetime  metric has an 
opposite signature  to theirs, $(-+++)$.
For definiteness, we recall  the conventional  notations for the two-point and three-point integrals,
\begin{eqnarray}
{i\pi^2\over {(2\pi )^4}} 
[B_0, -p_\mu B_1 ]=
\int_Q { [1, Q_\mu ] \over  (-Q^2+m_1^2) (-(Q-p)^2
+m_2^2)   } ,
\label{eq9p}
\end{eqnarray}
\begin{eqnarray}
&{i\pi^2\over {(2\pi )^4}}&
[C_0,\  -p_\mu C_{11}-p'_{\mu} C_{12} ,  \ 
p_\mu p_\nu C_{21}+ p'_\nu p'_\mu C_{22} +(p_\mu p'_\nu + 
p_\nu p'_\mu ) C_{23}
\   -g_{\mu \nu } C_{24} ] \cr 
&=& \int_Q  {  [1,Q_\mu ,  Q_\mu Q_\nu ] \over 
(-Q^2+m_1^2) (-(Q-p)^2+m_2^2)  (-(Q-p-p')^2
+m_3^2)  } ,
\label{eq9pp}
\end{eqnarray}
where the arguments for the $B- $ and $C-$ functions are defined as: 
$B_A (-p,m_1,m_2),  \  \ [A=0,1] $ and $   \  
C_B (-p, -p',m_1,m_2,m_3),  \ \  [B=0,11,12, 21,22,23,24] $.
In the algebraic derivation of the  one-loop amplitudes, 
we find it convenient to introduce the  definitions: 
$ p_\mu =\ud P_\mu +\rho_\mu , \ p'_\mu =\ud P_\mu -\rho_\mu , $ where $P=p+p', \ 
\rho = \ud (p-p')$.  The terms 
proportional to the Lorentz covariant $P^\mu = (p+p')^\mu $   will then  reduce, for the  full
Z-boson exchange amplitude in eq. (\ref{eq2}),  to negligible mass terms 
in the initial leptons. 

Dropping mass terms for all external fermions, the tensorial couplings cancel out and 
we need keep track of the vector couplings only, with the result:
\begin{eqnarray}
A_L^{JJ'}({\cal L}) &=& N_c\FFL \bigg [ a (f_L) m_f^2 C_0 
+a(f_R)\bigg (B_0^{(1)}-2C_{24}-m_{\tilde {f}'}^2C_0\bigg  )
+2a(\tilde {f}'_L) \tilde C_{24} + a(f_L) B_1^{(2)} \bigg ], \cr
A_R^{JJ'}({\cal L})&= &0. 
\label{eq9}
\end{eqnarray}
The cancellation of the right chirality vertex function  in this case 
is the reason behind our naming these contributions as ${\cal L}$-type.
The  two-point and three-point integrals functions  without 
a tilde symbol arise through the  fermion current  triangle contribution 
and  the self-energy contribution (represented by the term proportional 
to $ B_1^{(2)}$). These
 involve the argument variables  according to the following conventions, 
\hfill $B_A^{(1)}= B_A(-p-p', m_f, m_f), \ B_A^{(2)}= B_A(-p , m_f, m_{\tilde f'}),
\ B_A^{(3)}= B_A(-p', m_{\tilde f'}, m_f),  $ and 
$C_B = C_A (-p,-p',m_f,m_{\tilde f'} , m_f) .$
The integral functions  with a tilde arise in 
the  sfermion  current  diagram and are described by 
the argument variables, 
$\tilde C_A =  C_A (-p,-p',m_{\tilde f'}, m_f,m_{\tilde f'}).$ 

A very useful check on the above  results
concerns  the cancellation of ultraviolet divergencies. This
is indeed expected on the basis of the  general rule
that those interaction terms which are absent
from the classical  action, as is the case for the  flavor changing currents,
 cannot undergo renormalization. A  detailed discussion of this property
 is developed in \cite{soares}.  The  logarithmically divergent 
terms in eq.(\ref{eq9}), proportional to the quantity, 
$\D =-{2\over D-4}+\g -\ln \pi  $, as defined in \cite{pasvel}, 
arise from the two- and three-point integrals as, 
$ B_0 \to \D ,\ B_1 \to -\ud \D , \ C_{24}\to {1\over 4} \D  $, all other integrals  being finite. 
Performing these substitutions,  we indeed find that $\D $ comes accompanied  
by the overall factors, $ [-a(e_L)+a(\tilde {u_L^\star }) +a(d_R)], $ 
or $ [-a(e_L)+a(\tilde {d_R}) +a_R(u^c)], $ 
which  both  do  vanish  in the relevant configurations for $f, \tilde f'$.

Let us now consider the R parity odd Yukawa interactions 
involving the  $\l_{ijk}$. These contribute through the same triangle
diagrams as in Fig. \ref{fig2}. There arise
contributions of   ${\cal L}$-type,  in  the 
single configuration, $ f= e_k, 
\tilde f'=\tilde \nu^\star _{iL}$ and of 
${\cal R}$-type in the two  configurations, 
$f= {e_j\choose \nu_i} , 
\tilde f'={\tilde \nu_{iL} \choose \tilde e_{jL} }$. 
Following the same derivation as above, and neglecting  all of the external mass terms,  
we obtain  the  following results for the one-loop vector coupling vertex functions:  
\begin{eqnarray}
A_L^{JJ'}({\cal L})&=&  \FFIL [ a (f_L) m_f^2 C_0 
+a(f_R)(B_0^{(1)}-2C_{24}-m_{\tilde {f}'}^2 C_0 ) 
+2a(\tilde {f }'_L) \tilde C_{24} + a(f_L) B_1^{(2)} ], \cr
A_R^{JJ'}({\cal R})&= &\FF2L [ a (f_R) m_f^2 C_0 
+a(f_L)(B_0^{(1)}-2C_{24}-m_{\tilde {f}'}^2 C_0 )
+2a(\tilde {f}'_L) \tilde C_{24} + a(f_R) B_1^{(2)} ],  \cr 
&&
\label{eq10}
\end{eqnarray}
with $A_R^{JJ'}({\cal L})= 0, \  A_L^{JJ'}({\cal R})= 0.$
We note that the ${\cal {L} , \ {R} } $ contributions 
are related by a mere  chirality flip transformation and that the color 
factor, $N_c$, is absent in the present case.
\subsection{Down-quark-antiquark pairs}
\label{subsec2}
The processes involving  flavor  non-diagonal 
final down-quark-antiquark pairs, $l^-(k)+l^+(k')\to  d_J(p)+\bar d_{J'}(p')$,
pick up non vanishing contributions  only from  the $\l '_{ijk}$
interactions.  Our discussion here will be brief 
since this case is formally
similar to the leptonic case treated  in subsection \ref{subsec1}. 
In particular, the external fermions  masses, 
 for all three families,  can be neglected 
to a good approximation at the energy scales of interest.
The tree  level amplitude comprises an ${\cal R}$-type 
single t-channel  $\tilde u$-squark 
exchange diagram and  two   s-channel diagrams involving 
 $\tilde \nu  $ and   $\tilde  {\bar \nu  } $ sneutrinos
of the type shown in $(a) $ of  Fig. \ref{fig1},
\begin{eqnarray}
M^{JJ'}_t&=& -   
{ \l '_{1jJ} \l ^{'\star}_{1jJ'}\over 2(t-m^2_{\tilde u_{jL} }) } 
\bar u_R(p) \g_\mu v_R(p')\bar v_L(k')\g_\mu u_L(k) \cr
&-& {1 \over s-m^2_{\tilde \nu_{iL} } }\bigg [\l _{i11}\l ^{'\star } _{iJJ'}
\bar v_R(k') u_L(k) \bar u_L(p) v_R(p') +
\l _{i11}^\star \l ' _{iJJ'}
\bar v_L(k') u_R(k) \bar u_R(p) v_L(p')\bigg ] ,\cr 
&&
\label{eqp1}
\end{eqnarray}
where a  Kronecker symbol factor, $\d_{ab} $,  expressing the dependence
on the  final state quarks color indices, $d^a \bar d_b $,  
has been suppressed. This dependence will induce in 
the analog of the formula in  eq.(\ref{eqr4}) expressing  the rates,
an extra color factor,  $N_c$. 

At one-loop level, 
the  dressed   $Z\  d_J\  \bar d_{J'} $ vertex functions in the   
Z-boson s-channel exchange amplitude can be described by the 
same type of triangle diagrams as in Fig. \ref{fig2}. 
The fields configurations circulating in the loop
correspond now  to  quarks-sleptons of ${\cal L}$-type,
$ f=d_k;  \tilde l'= \tilde \nu^\star _{iL}, $ and of ${\cal R}$-type,  
$ f={ d_j \choose u_j}; \ \tilde l'= 
{\tilde \nu _{iL} \choose \tilde e_{iL} }$. There also  occurs
corresponding leptons-squarks fields  configurations  of ${\cal L}$-type,  
$ l=\nu_i^c  ; \tilde f'= \tilde d_{kR}$, and ${\cal R}-$type, 
$ l= {\nu_i \choose e_i};  \tilde f'= 
{\tilde d_{jL} \choose \tilde u_{jL} }$. 
The ${\cal L }-$ and ${\cal  R  }-$   type contributions
differ by a  chirality flip, the first contributing  to 
$A_L^{JJ'}$ and the second to $A_R^{JJ'}$.
The calculations are formally similar  to those in subsection \ref{subsec1} and 
the final results  have a nearly identical structure to those 
given in (\ref{eq9}). For clarity, we quote the final 
formulas  for the one-loop vector coupling vertex functions, 
\begin{eqnarray}
A_L^{JJ'}({\cal L})&= &\FFQ [ a (f_L) m_f^2 C_0 
+a(f_R)(B_0^{(1)}-2C_{24}-m_{\tilde {f}'}^2 C_0 ) 
+2a(\tilde {f}'_L) \tilde C_{24} + a(d_L) B_1^{(2)} ], \cr
A_R^{JJ'}({\cal R})&= &\FFP [ a (f_R) m_f^2 C_0 
+a(f_L)(B_0^{(1)}-2C_{24}-m_{\tilde {f}'}^2 C_0 )
+2a(\tilde {f}'_L) \tilde C_{24} + a(d_R) B_1^{(2)} ],  \cr 
&&
\label{eq10p}
\end{eqnarray}
where the intermediate fermion-sfermion fields are labelled  by the 
indices $ f, \ \tilde f' $.  There are implicit sums
 in eq.(\ref{eq10p})   over the 
above quoted leptons-squarks and  quarks-sleptons 
configurations. The attendant ultraviolet divergencies  are accompanied 
 again with vanishing factors, $ a(\tilde d_R) -a(d_L)+a(\nu^c_R) =0,  \  \ 
 a(\tilde d_L) -a(d_R)+a(\nu_L) =0.$ 

\subsection{Up-quark-antiquark pairs}
\label{subsec3}
 The production processes of  
up-quark-antiquark pairs of different families, 
$l^-(k)+l^+(k')\to u_J(p)+\bar u_{J'}(p')$,  
may be  controlled by the $\l'_{ijk}$ interactions only.
The tree amplitude is associated with an u-channel 
$\tilde d$-squark exchange,  of  type  similar to that
 shown by $(a)$ in  Fig. \ref{fig1}, and  can be expressed  as,
\begin{eqnarray}
M^{JJ'}_{t}= -{\l ^{'\star }_{1Jk} \l '_{1J'k}
\over 2(u-m^2_{\tilde d_{kR} }) } 
\bar v_L(k') \g^\mu u_L(k)\bar u_L(p)\g_\mu v_L(p'),
\label{eq1pp}
\end{eqnarray}
after using the   Fierz reordering identity, appropriate to commuting 
Dirac (rather than  anticommuting field)
spinors, $\bar u^c(k) P_L v(p')  \bar u(p) P_R v^c(k')  =+\ud  
\bar v_L(k') \g^\mu u_L(k)\bar u_L(p)\g_\mu v_L(p') .$ 
We have  omitted the  Kronecker symbol
$\d_{a b} $ on the $u^a \bar u_b$ color indices, which  will 
result in an extra  color factor $N_c=3$  
for the rates, as shown  explicitly in eq.(\ref{equ4}) below.
The present case is formally similar to the leptonic case treated 
in subsection \ref{subsec1}, except for  a chirality flip in the final fermions.
We are especially interested 
here in final states containing a top-quark, such as $t\bar c$ or $t \bar u$,
for which external particles mass terms cannot obviously  be ignored. 
The  equation, analogous to (\ref{eqr4}), which  expresses the 
summations over the initial and final  polarizations 
in the total  (tree and loop) amplitude, 
takes now  the form, 
\begin{eqnarray}
\sum_{pol} \vert M^{JJ'}_t+M^{JJ'}_l\vert^2&=&  N_c\bigg [ \bigg 
\vert -{  \l'_{1J'k} \l {'^\star } _{1Jk} \over 
2(u-m^2_{\tilde d_{kR} } )}  +\bigg ({g\over 2\cos \t_W}\bigg )^2
{  a(e_L) A_L^{JJ'}(u,s+i\e ) \over 
s-m_Z^2+im_Z\G_Z   } \bigg \vert^2 \cr &\times & 
16 (k\cdot p')(k'\cdot p ) + 8m_Jm_{J'} 
(k\cdot k') \varphi_{LR} ({\cal L})  \bigg ]  \ , 
\label{equ4}
\end{eqnarray}
where  $O(m_e^2)$ terms were ignored and  we  have denoted,
\begin{eqnarray}
\varphi_{LR}({\cal L}) &=& +\bigg ({g\over 2\cos \t_W}\bigg )^2
\bigg ({ a(e_L) A_{R}^{JJ'}(i\e ) \over 
s-m_Z^2+im_Z\G_Z }\bigg )^\star  \bigg ({ \l'_{1J'k}\l _{1Jk}^{'\star }\over 
2(u-m^2_{\tilde d_{kR} } )  } \bigg ) +c.\ c\ . 
\label{equ4p}
\end{eqnarray}
The modified  structure for the kinematical factors  in the above 
up-quarks case, eq.(\ref{equ4}),  in comparison with the leptons  and d-quarks case, eq.(\ref{eqr4}),  reflects the  difference in 
chiral structure for the RPV  tree  level amplitude. 

In the massless limit  for both  the initial and final fermions 
(where helicity, $h=(-1,+1)$, and chirality, $H=(L,R)$, coincide) 
the RPV interactions contribute to the helicity amplitudes for the  process, 
$l^-+l^+\to f_J +\bar f_{J'}$, in  the mixed type helicity configurations, 
 $ h_{l^-} =-h_{l^+},  \ h_{f_J} =-h_{\bar f_{J'} }$, (same as for the RPC 
gauge interactions)  which are further  restricted  by the conditions, 
$h_{l^-}= -h_{f_J}$, for leptons and d-quarks production, and
 $h_{l^-}= h_{f_J}$, for up-quarks production.   The dependence of the  RPV scattering
amplitudes on scattering angle has a kinematical
factor in the numerator of the form,  $[ 1+h_{l^-} h_{f_J} \cos \theta ] ^2.$ 
[The parts in our formulas in 
eqs. (\ref{equ4}) and (\ref{eqr4}), containing the interference terms between RPV and RPC contributions,  partially agree with the published results \cite{kalin12,kalin13}. We disagree
with \cite{kalin12,kalin13} on the relative signs of RPV and RPC contributions 
and  with \cite{kalin13} on the helicity structure for the up-quarks case.
Concerning the latter up-quarks case, our results concur with those 
reported in a recent study \cite{mahanta}.]

The states in the internal loops of the triangle diagrams  occur 
in two distinct ${\cal L}$-type  configurations, 
$ f={d_k  \choose e_i^c} ; \ \ \tilde f'= 
{\tilde e_{iL}^\star  \choose   \tilde d_{kR}}$. 
The calculations involved in keeping track of 
the mass terms are rather tedious.  They
were performed by means of the mathematica software package, 
``Tracer" \cite{jamin} whose results   were checked against those obtained 
by means of  ``FeynCalc" \cite{calc}.
The relevant formulas for the vertex functions  read:
\begin{eqnarray}
 A_L^{JJ'}({\cal L})& =&
 \FLU \bigg [  {a_L(u) }\, {B_1^{(2)}} + 
      {a(f_L)}\, {{ {m_f}}^2}  C_0 + 
      {a(\tilde  f')}\bigg (    
     2 {\tilde C_{24}} + 
      2 \   m_J^2\, ( {\tilde C_{12}}\, - 
      {\tilde C_{21}}\,  + 
      {\tilde C_{23}}\, - 
      {\tilde C_{11}}\, ) \bigg )  \cr &+& 
         {a(f_R)}\, \bigg ( 
       {B_0^{(1)}} - 2\, {C_{24}} - 
         {{ {m_{\tilde  f'}}}^2} {C_0}  + 
         {{ {m_{J}}}^2}   
         \bigg ( C_0 +3 C_{11} -2C_{12} +2C_{21}
	 -2C_{23}\bigg )   -m_{J'}^2 C_{12} \bigg ) \bigg ], \cr 
 A_R^{JJ'}({\cal L})& = & \FLU  m_J m_{J'}  \bigg [ 
      2 {a(\tilde  f')}\,\bigg  ( - {\tilde C_{23}} +  {\tilde C_{22}}\bigg ) + 
  {a(f_R)}\,
        \bigg  (  -C_{11} +C_{12} -2C_{23} +2 {C_{22}} \bigg )
 \, \bigg ] . \cr
 &&
\label{eqm11}
\end{eqnarray}
The above formulas include an implicit sum over the  
two allowed configurations for the  internal fermion-sfermions, namely,
$a(d_ {kH}), \   a(\tilde e^\star _{iL} ) $     and 
$a(e_{iH}^c ), \ a(\tilde d_{kR} ) $.
For completeness, we also display the formula  expressing the tensorial covariants,
\begin{eqnarray}
&&\G_\mu ^Z (p,p')_{tensorial}= \FLU i\s_{\mu \nu }p^\nu \bigg [ m_J P_L \bigg (
a(f_R) (C_{11}-C_{12}+C_{21}-C_{23}) \cr &-&
a(\tilde f ') (\tilde C_{11}+\tilde C_{21}-\tilde C_{12}-\tilde C_{23})
\bigg )  
 + m_{J'} P_R \bigg (
+a(f_R) (C_{22}-C_{23})
+a(\tilde f ') (\tilde C_{23}-\tilde C_{22}) \bigg ) \bigg ]. \cr
&&
\label{eqm11t}
\end{eqnarray}
The complete  $Z f_J\bar f_{J'}$ vertex function, $\G^\mu = \G_{vectorial}^\mu  + 
\G_{tensorial}^\mu $, should (after extracting the external Dirac spinors and the RPV coupling
constant  factors) be symmetrical under the interchange, $J\leftrightarrow J'$, 
or more specifically,  under the interchange,
$m_J\leftrightarrow  m_{J'}$. This property is not explicit on the 
expressions in eqs. (\ref{eqm11}) and (\ref{eqm11t}), but can be established by 
reexpressing the Lorentz covariants by means of the Gordon identity. The
naive use of eq.(\ref{eq6p}) to compute CP-odd asymmetries would seem to 
yield finite contributions  (even in the absence of a CP-odd phase) from the mass terms 
in the vectorial vertex functions, $A_{L}^{JJ'}$,  
owing to their lack of symmetry under, $m_J\leftrightarrow m_{J'}$. 
Clearly, this cannot hold true and is an artefact of restricting to the vectorial couplings.
Including the tensorial couplings is necessary  for a consistent  treatment
of the contributions depending on the external  fermions masses.
Nevertheless, we emphasize that the tensorial vertex  contributions will not included in 
our numerical results.

Finally, we add a general comment  concerning the photon vertex functions,
$ A_{L,R}^{\g JJ'} $,  and the way to incorporate the 
$\g $-exchange contributions in the total
amplitudes, eqs.(\ref{eqr4}) and (\ref{equ4}).
One needs to add terms obtained by  substituting,
${g\over 2 \cos \t _W } \to {e\over 2} = { g \sin \t_W \over 2},
\ a_{L,R}(f)\to 2Q(f), \ (s-m_Z^2  +im_Z \G_Z)^{-1}\to s^{-1},$ along
with the substitution of Z-boson by photon vertex functions,
$ A_{L,R}^{JJ'} (\tilde e, s+i\e ) \to A_{L,R}^{\g JJ'} (\tilde e, s+i\e )$.
The substitution which adds in both  Z-boson and photon exchange 
contributions reads explicitly:
$$ [a_{R,L}(e) A^{JJ'}_{L,R} ]\to
\bigg [ a_{R,L}(e) \sum_f a(f) C_f+ 2Q(e)  \sin^2 \t_W \cos^2 \t_W
[(s-m_Z^2+im_Z \G_Z) / s] \sum_f 2Q(f) C_f   \bigg ]  , $$
where we have used the schematic representation, 
$A^{JJ'}_{L,R}=\sum_f a(f) C_f$.
\section{ Basic assumptions and results}
\label{sec5}
\subsection{General  context of flavor changing physics}
To place the discussion of the RPV  effects  in perspective, we  
briefly review the current situation of flavor changing physics.
In the standard model, non-diagonal effects with respect 
to  the quarks flavor arise through loop
diagrams. The typical structure of one-loop contributions to, say,
 the $Zf\bar f$ vertex function, $\sum_i
V^\star _{iJ} V_{iJ'}  I(m_i^{f2}/m_Z^2) $, involves   a
summation over quark families of
CKM matrices factors times a 
loop integral. This schematic  formula shows explicitly how 
the CKM matrix unitarity,  along with the near quarks  masses  degeneracies 
relative to the Z-boson mass scale 
(valid for all quarks with the  exception of the top-quark) strongly
suppresses flavor changing  effects. Indeed, for the  down-quark-antiquark case,
the Z-boson decays branching fractions, $B_{JJ'}$, 
 were estimated at the values, 
 $10^{-7}$ for $(\bar b s + \bar s b)$, $10^{-9}$ for $(\bar b d+\bar d b) $, 
 $10^{-11}$ for $(\bar s d+\bar d s)$,    and the  corresponding 
CP  asymmetries, ${\cal A}_{JJ'}$, at  the values,
$ [10^{-5}\ , 10^{-3}, \  10^{-1}]  \sin \d_{CKM} $
\cite{bernabeu,hou}, respectively. 

By contrast,  flavor changing effects are  expected to  attain
observable levels in several  extensions of the standard model.  
Thus,  one to three order of magnitudes can be  gained on rates $B_{JJ'}$ 
in models accommodating a  fourth  quark family \cite{bernabeu,hou}.
For the  two Higgs doublets extended standard model,   a recent comprehensive study of
fermion-antifermion  pair production at 
leptonic colliders \cite{atwood}  quotes 
for the  flavor changing rates,  $B_{JJ'}\approx 10^{-6} - 10^{-8} $ for 
$Z\to (\bar b +s) +(\bar s +b)$ and    $\s_{JJ'}\approx 10^{-5} - 10^{-6} R$, where,  
$ R =\s (e^++e^- \to \mu^++\mu ^-)=  4\pi \a^2/(3s)= 86.8 /(\sqrt s )^ 2 
fbarns \ (TeV)^{-2}$. Large CP violation signals  are also found in the 
reaction, $p\bar p \to t\bar b X$,  in  the two Higgs doublets and supersymmetric 
models \cite{atwood1}.
 
For  the minimal supersymmetric standard model, 
due to the  expected nearness of
superpartners  masses to $m_Z$,
flavor changing  loop corrections 
can become  threateningly large,  unless 
their contributions are bounded  by  postulating either  a degeneracy of the 
soft supersymmetry breaking scalars masses parameters 
 or  an alignment of the  fermion and  scalar 
superpartners current-mass bases transformation matrices.
An early calculation of the  contribution to  Z-boson decay
 flavor changing rates,   $Z\to q_J \bar q_{J'} $, induced by  radiative corrections from gluino-squark triangle diagrams of squarks flavor mixing, 
found \cite{duncan}:
 $B_{JJ'} \approx 10^{-5}$.
This result is suspect  since a more complete calculation of the effect 
performed subsequently \cite{mukho} obtained considerably smaller 
contributions.   Both calculations rely  on  unrealistic inputs, 
including a wrong mass for the top-quark and too low values for 
the  superpartners mass 
parameters. It is hoped that  a complete  updated study 
could be soon performed. 
In fact, during the last few years, the study  of loop corrections in 
extended versions of the standard model has evolved into  a streamlined activity.
For instance, calculations of loop contributions to 
the magnetic moment of the  $\tau $-lepton or of 
the heavy quarks, such as those reported in \cite{bernabeu2} (two-Higgs doublets model) or in 
\cite{hollik} (minimal supersymmetric standard model) could be usefully transposed to the case  
of fermion pair production observables.

The  information from experimental searches  on flavor changing 
physics at high energy colliders is rather meager \cite{branch1}. 
Upper bounds for  the leptonic Z-boson branching ratios, $B_{JJ'}$, 
are reported \cite{branch2} at, 
$ 1.7 \ 10^{-6}$ for $( \bar e \mu +\bar \mu e) $,  $9.8 \ 10^{-6} $ for 
$( \bar e \tau +\bar \tau e) $ and $  1.7 \  10^{-5} $ for $(\bar \mu \tau +\bar \tau  \mu )$. 
No results have been quoted  so far   for  $d-$ or $u-$ quark 
pairs production, reflecting the hard experimental problems faced in 
identifying quarks flavors at high energies.   The prospect  for experimental measurements
at the future leptonic colliders is  brightest
 for cases involving one top-quark owing  
to the easier  kinematical identification offered by the large  mass 
disparity  in  the final  state jets.
For leptonic colliders at energies above those of LEP,
the  reactions involving the 
production of  Higgs or heavy $Z'$-gauge  bosons 
which subsequently decay to fermion pairs
could be effective   sources  of flavor non-diagonal
effects, especially when a 
top-quark is produced. At still higher energies, in the TeV regime, the 
production subprocesses involving collisions  of gauge bosons  pairs 
radiated by the incident leptons, as in  $l^-+l^+ \to W^-+W^++\nu  +\bar \nu $, 
could lead to  flavor non-diagonal final states, such as,
 $\nu +\bar \nu + t + \bar c $  with rates of order a few  fbarns \cite{hou2}.

\subsection{Choices of parameters and models}
Our main  assumption in this work  is that
no other sources besides  the R parity odd interactions 
contribute significantly to the flavor changing rates and CP asymmetries.  
However, to infer useful information from possible  future  experimental 
results, we must deal with two main types of 
uncertainties. The first concerns the family structure of 
the coupling constants.  On this issue,   one can only 
postulate specific hypotheses or make  model-dependent statements. 
At this point, we may note that the experimental 
indirect  upper bounds  on single coupling constants  are typically, 
 $\l < 0.05  $ or $ \l '  < 0.05$ times ${\tilde m\over 100 GeV}$, 
 except for three special  cases where strong bounds exist: 
 $\l'_{111} < 3.9 \ 10^{-4} ({\tilde m_q\over 100 GeV} )^2
 ({\tilde m_g\over 100 GeV} )^\ud , \quad  (0\nu \b \b -$  decay \cite{hirsch}) 
 $\l '_{133}< 2 \ 10^{-3} \ (\nu_e $ mass \cite{godbole}) and 
 $\l'_{imk} <  2. \ 10^{-2} ({ m_{\tilde d_{kR}  } \over 100 GeV} ), 
 \ [i, \ k =1,2,3; \ m=1,2]  $, ($K\to \pi \nu \bar \nu $ \cite{agashe}).
Strong  bounds  have  been derived
 for products of coupling constants pairs in specific
 family configurations. For instance,  a valuable  source 
 for  the $\l_{ijk}$ coupling constants is provided by the rare decays, $e^-_l\to e^-_m+
e^-_n+e^+_p$ \cite{roy}, which  probe the combinations of coupling constants, $F_{abcd} = \sum_i
({100 GeV \over m_{\tilde \nu_{iL} } })^2 
\l_{iab} \l ^\star _{icd}$. Except for the  strong bound,
$F_{1112}^2 + F^2_{2111}< 4.3 \ 10^{-13},\   [\mu \to 3e]  $
 the other combinations  of coupling constants involving the third generation are less 
strongly bound, as for instance, 
$ F_{1113}^2 + F^2_{3111}< 3.1 \ 10^{-5} \ [\tau  \to 3e]$
\cite{roy}.  
 Another useful source   is provided by the
neutrinoless double beta decay  process \cite{babu,hirsch,hirsch1}.
The strongest bounds occur  for the following configurations of  flavour indices
(using the reference value ${\tilde m = 100 GeV} $): 
$\l'_{113}\l'_{131}< 7.9 \times 10^{-8}, \ 
\l'_{112}\l'_{121}< 2.3 \times 10^{-6}, \ \l ^{'2}_{111}< 4.6 \times 10^{-5} $,
quoting from \cite{hirsch1} where
the  initial analysis of \cite{babu} was updated. 
Finally,  the strongest bounds deduced from  neutral mesons  $(B\bar B, \ K\bar K)$ mixing parameters are:
$ F'_{1311}< 2\ 10^{-5}, F'_{1331}< 3.3\ 10^{-8}, \ 
F'_{1221}< 4.5\ 10^{-9},$ \cite{roy}, where 
$F'_{abcd} = \sum_i ({100 GeV \over m_{\tilde \nu_{iL} } })^2 
\l '_{iab} \l ^{'\star }_{icd}$.

The second type  of uncertainties concerns the spectrum of scalar superpartners.
At one extreme, are  the experimental lower bounds, which   reach   for sleptons, 
$40-65 $ GeV,  and for   squarks, $90 - 200$ GeV,  and at the other extreme,  
the theoretical naturalness requirement   which sets
an  upper bound at $1 $ TeV.

In order to estimate  the uncertainties on predictions emanating from 
the above  two sources, it is necessary to delineate the dependence of amplitudes  on  sfermion masses.
Examining the structure of the  relevant  contributions to flavor changing rates
for, say,  the lepton case,  we note 
that the t-channel exchange   tree amplitudes are given by  a onefold summation 
over sfermions  families, $\sum_i \vert t^i_{JJ'}\vert /\tilde m_i^2$, involving the  combination  of coupling constants,  $ t^i_{JJ'}=   \l _{i1J} \l_{i1J'}^\star $. 
The typical structure  for the leptonic loop amplitudes is a twofold summation
over fermions and sfermions families,
$\sum_{ij}  l^{ij}_{JJ'} F_{JJ'}^{ij} (s+i\e ) , \ $ where $ l^{ij}_{JJ'}=  \l_{ijJ}
\l^\star _{ijJ'} $, and the   loop integrals, $F^{ij}_{JJ'} $,  have a  non-trivial
dependence on the fermions and sfermions masses, as exhibited 
on the formulas  derived in subsections \ref{subsec1},
\ref{subsec2} and \ref{subsec3} [see, e.g., eq.(\ref{eq10p})].

The effective dependence on the superparticle masses 
involves  ratios, $m_f^2/\tilde m^2 $ or $s/\tilde m^2 $, in such a way that 
the dependence is suppressed for large $\tilde m$. 
(Obviously, $s= m_Z^2$  for Z-boson pole observables.)
In  applications such as ours where,     $s\ge m_Z^2$,
all the  fermions, with the exception of the top-quark,   can be  regarded as 
being massless.
In particular, the  first two light families (for either $  l, \  d,\  u $)
should have  comparable contributions, the third 
family behaving most  distinctly in the  top-quark case. A quick analysis,  
taking the explicit mass factors into account, indicates that  
 loop amplitudes should  scale with sfermions masses as, $(s/\tilde m^2)^n$, with 
a variable  exponent ranging in the interval,  $1< n< 2$. 
Any possible enhancement effect from  the explicit sfermions 
mass factors in  eq.(\ref{eq10p}) is moderated 
in the full result by  the fact that 
the accompanying loop integral factor  has itself
a power decrease with increasing  $\tilde m^2$.
Thus, the Z-boson pole rates should depend on the masses $\tilde m$  roughly 
as $(1/\tilde m^2)^{2n}$, while the off Z-boson  pole rates, being determined 
by the tree amplitudes, should behave  more nearly as $(1/\tilde m^2)^2$.
As for the asymmetries, since these are given by ratios of squared amplitudes, 
one expects them to have a  weak sensitivity on the  sfermion
masses.

To infer   the physical implications on the RPV coupling constants, 
we avoid making too   detailed model-dependent 
assumptions on the scalar superpartners spectrum.
Thus, we shall neglect mass splittings 
between all the sfermions and   set uniformly  all   
 sleptons, sneutrinos and squarks masses
at a unique  family (species)  independent value, $\tilde m$, 
chosen to vary in  the wide  variation interval, 
$100 < \tilde m <  1000$ GeV. This prescription should suffice for the kind of 
semi-realistic  predictions at which we are aiming. This approximation
makes more transparent the dependence on the RPV 
coupling constants, which then involves the 
quadratic products designated by $t^i_{JJ'}$ (tree)  and $l^{ij}_{JJ'}$ (loop),
where the dummy family indices refer to sfermions (tree)
and fermion-sfermions (loop).
For  off Z-boson-pole observables,  flavor non diagonal
rates  are controlled by products of  two  
different couplings, $\vert t_{JJ'}^{i} \vert ^2 ,$ and  asymmetries  by 
normalized products of four different couplings, $Im(  t^{i' \star }_{JJ'} l_{JJ'}^{ij} ) / \vert t_{JJ'}^{i''} \vert ^2.$
For Z-boson pole observables,  rates and asymmetries 
are again  controlled by  products of two and
 four different coupling constants, 
$\vert l_{JJ'}^{ij} \vert ^2 $  and  $Im(  l^{i'j' \star }_{JJ'} l_{JJ'}^{ij} ) 
/ \vert l_{JJ'}^{i'' j''} \vert ^2,$  respectively.  Let us note that  if
the off-diagonal rates were dominated by some alternative mechanism, 
the asymmetries  would then  involve products
of four different coupling constants rather than the above ratio. 
 
It is useful here to  set up  a catalog of the 
species and families   configurations for the 
sfermions (tree)  or fermion-sfermions (loop)  involved in the various cases.
 In the tree level amplitudes, these configurations are  for leptons: 
$t^i_{JJ'}=  \l_{iJ1}\l ^\star _{iJ'1},\ \tilde \nu_{iL}  \  ({\cal L}$-type),
$t^i_{JJ'}=  \l_{i1J} \l ^\star _{i1J'},\ \tilde \nu _{iL}  \  ({\cal R}$-type);
for d-quarks,  
$ t_{JJ'}^j = \l ' _{1jJ}\l ^{'\star } _{1jJ'},   \ \tilde u_{jL}\  ({\cal R}$-type);
for u-quarks, 
$t_{JJ'}^k= \l ^{'\star }_{1Jk}\l ' _{1J'k},  \  \tilde d_{kR} \  ({\cal L}$-type).
In the loop level  amplitudes,  the  
coupling constants and internal fermion-sfermion configurations  are for leptons: 

$l^{jk}_{JJ'}= \l ^{'\star }_{Jjk} \l '_{J'jk}, \ 
[{d_k \choose \tilde u^\star _{jL}  } ,
\ {u _j^c \choose \tilde d_{kR}}];  \quad 
l^{ik}_{JJ'}= \l ^{\star }_{iJk} \l _{iJ'k}, \ 
{e_k \choose \tilde \nu^\star _{iL}  } \ 
({\cal L}$-type);    

$l^{ij}_{JJ'}= \l _{ijJ} \l ^{\star }_{ijJ'}, \ 
[{e_j \choose {\tilde  \nu  }_{iL} } ,
\ [{\nu _i \choose \tilde e_{jL}}]  \  ({\cal R}$-type);

for d-quarks:

$l^{ik} _{JJ'}=\l ^{'\star }_{iJk} \l ' _{iJ'k}, \ 
[{d_k \choose \tilde \nu^\star _{iL}} ,
\ {\nu^c_i \choose \tilde d_{kR}}]  \  ({\cal L}$-type);

$l^{ij} _{JJ'}=\l ^{'\star }_{ijJ'} \l ' _{ijJ}, \ 
[{d_j \choose \tilde \nu_{iL}} , \ {u_j \choose \tilde e_{iL}}; 
\quad {\nu_i \choose \tilde d_{jL}} , \ {e_i \choose \tilde u_{jL}}] \   ({\cal R}$-type);

 for u-quarks,

$l_{JJ'}^{ik} = \l ^{'\star }_{iJk} \l'_{iJ'k}, \ 
[{d_k \choose \tilde e^\star _{iL}} , \ {e^c_i \choose \tilde d_{kR}}]  \ 
({\cal L}$-type).

We shall present numerical results for a subset of the above list of cases. 
For leptons and d-quarks, we  shall restrict  consideration 
to the ${\cal R}$-type  terms which contribute to the Z-boson  vertex 
function,  $A_R$.  We also retain the sleptons-quarks internal states for 
d-quark production  (involving $\l^ {'\star }_{ijJ'} \l '_{ijJ}$) 
and the sleptons-leptons for lepton  production  
(involving $\l^ {\star }_{ijJ'} \l _{ijJ}$) 
For the up-quark production, we consider the ${\cal L}$-type terms
(involving $\l^ {'\star }_{iJk} \l '_{iJ'k}$) and, for the  off Z-boson pole
case, omit the term $\varphi_{LR} $ in eq.(\ref{equ4}) in the numerical results.

Since  the running family index in the  parameters relevant to  tree level amplitudes refers to sfermions,  
consistently with the approximation of a uniform family independent
mass spectrum, we may as well consider that index as being fixed.
Accordingly, we  shall set  
these parameters at the reference value,  $t^i_{JJ'}=10^{-2}$.  
In contrast to the off Z-boson pole  rates, 
the asymmetries depend non trivially on the  fermion mass spectrum through  
one of the two family indices  in $l^{ij}_{JJ'}$  ($i $ or $j$)
associated to fermions. To discuss our 
predictions, rather than going through the list of four distinct coupling 
constants, we shall make certain general hypotheses regarding the generation 
dependence of the RPV interactions for the fermionic index.
At one extreme is the case where all three 
generations are treated alike, the other extreme being
the case where only one generation 
dominates. We shall  consider three different 
cases which are distinguished  by the interval
over which the fermions family indices are 
allowed  to range in the quantities,  $l_{JJ'}^{ij}$. We define  Case {\bf A } by the  prescription  of  equal values for
all three families of fermions ($i=1,2,3$); Case {\bf  B},  for
the second and third  families ($i=2,3$);
and  Case {\bf C},  for the third family only ($i=3$).
For all three cases, we set the
relevant parameters uniformly  at the reference values, 
$l_{JJ'}^{ij}=10^{-2}$. While the   results in Case {\bf C} reflect directly 
on the situation associated with the 
hypothesis of dominant third family configurations, the corresponding 
results in situations where the  first or second family  are assumed dominant, 
can be deduced by taking the differences between the results in 
Cases {\bf A} and {\bf B} and  Cases {\bf B} and {\bf C}, respectively.

In order  to obtain non-vanishing 
CP asymmetries, we still  need to specify   a prescription 
for introducing a relative 
CP-odd complex  phase, denoted  $\psi $, between the 
various RPV coupling constants. 
We shall set  this at the reference value,  $\psi =\pi /2$.
Since the CP asymmetries are proportional
to  the  imaginary part of the phase factor,
the  requisite dependence is simply 
reinstated by inserting the overall factor, $\sin \psi $.
Different prescriptions must be implemented  depending on
whether one considers observables at or off the Z-boson pole.
The Z-boson pole asymmetries are controlled   by a relative complex 
phase  between the  combinations of coupling constants denoted,
$l_{JJ'}^{ij}$ only. For definiteness, 
we choose here to assign a non-vanishing complex phase only 
to the third fermion family, namely, $arg(l^{ij}_{JJ'}) =[0,0,\pi /2]$,
for $[i \ or  \  j =1,2,3]$.  In fact, a relative phase between light families only 
would contribute insignificantly to the 
 Z-boson pole asymmetries,  because of the antisymmetry in $\a \to 
\a '$  in eq.(\ref{eq5}) and the fact that $F^{JJ'} _\a (m_Z^2)$   are  approximately  equal  when the fermion index in $\a =(i,j)$ belongs to the two first  families. 
The off Z-boson pole asymmetries  are controlled by a relative complex phase between the tree 
and loop level amplitudes. For definiteness,  we  choose here to 
assign a vanishing argument to the  coupling constants combination, $t^i_{JJ'}$ 
appearing at  tree level and   non-vanishing arguments to the full set of   
loop amplitude combinations, namely,  $arg(l_{JJ'}^{ij}) =\pi/2 , $  
where  the  fermion index ($i \ or \ j $  as the case may be) 
varies  over   the  ranges  relevant  to each of the three  cases 
{\bf A, \ B, \ C}.
 
\subsection{Numerical results and discussion}
\subsubsection{Z-boson decays observables}
We  start by presenting  the numerical results for the integrated 
rates associated with Z-boson decays into fermion pairs.  
These   are given 
for the d-quarks,   leptons and u-quarks cases 
in Table \ref{table1}.  We observe a fast decrease  of rates with increasing
values  of the  mass parameter, $\tilde m$.  Our results can be approximately   fitted by a  power law  dependence 
 which is intermediate between $\tilde m^{-2} $ 
and $\tilde m^{-3}$.  Explicitly, 
 the Z-boson flavor non diagonal   decay rates to d-quarks, leptons and u-quarks, 
are found to scale approximately 
as, $ B_{JJ'} \approx ({\l_{ijJ} \l_{ijJ'} \over 0.01})^2 
({100 GeV \over \tilde m})^{2.5} \times 10^{-9} [5. , \ 1., \ 2. ] $, respectively. 
When a top-quark  intermediate state is allowed  in the loop amplitude,
this dominates over the contributions from the light families. This is 
clearly seen on the  d-quarks results which are somewhat larger than  those for up-quarks  and  significantly larger 
 than  those for leptons, the more so for larger $\tilde m$.  This result is
explained partly by the color factor, partly by the
presence of the  top-quark contribution  only 
for the down-quarks case. For contributions involving other intermediate 
states than up-quarks,  whether the internal 
fermion generation index in the RPV coupling constants, $\l_{ijk}$, 
runs over all three generations (Case {\bf A}), 
the second and third generations (Case {\bf B}) 
or the third generation only (Case {\bf  C}),  we find  that   
rates get  reduced by factors  roughly less than 
2 in each  of these stages.  Therefore,  this comparison indicates  
a certain degree of family independence for  the Z-boson branching fractions 
for the cases where  either leptons or d-quarks propagate inside the loops. 
 
Proceeding next to the CP-odd asymmetries, since  these are proportional to
ratios of the  RPV coupling constants,  it follows 
in our prescription of  using uniform values for these, that asymmetries
must be independent of the specific reference value chosen. As for their 
dependence on $\tilde m$,  we  see on Table \ref{table1} 
that this is rather  strong and that the sense of variation with increasing 
$\tilde m$ corresponds (for absolute values of ${\cal A}_{JJ'}$) to a decrease 
for d-quarks and  an increase for u-quarks and leptons.
The comparison of different production cases 
shows that the CP asymmetries are largest, $O(10^{-1})$, for d-quarks at
small $\tilde m \simeq 100 GeV$, and for u-quarks at 
large $\tilde m \simeq 1000 GeV$.  For leptons, the asymmetries are systematically
small, $O(10^{-3} \ - \ 10^{-4})$. The above features are explained by the occurrence for
d-quarks production of an intermediate top-quark contribution  and also 
by the larger  values of the rates  at large $\tilde m$ in  this case.
The comparison of results in Cases {\bf A} and   {\bf B} indicates that the first
two light families give roughly equal contributions in all cases.

For Case {\bf C}, the CP-odd asymmetries are vanishingly small,
as expected from our prescription of assigning the CP-odd phase, since Case {\bf C}
corresponds then to a situation where only single pairs of coupling constants 
dominate. Recall that for  the specific cases considered in the numerical applications, 
namely, ${\cal R}$-type  for d-quarks and leptons and 
${\cal L}$-type  for u-quarks,  the relevant products of RPV coupling constants are, 
$\l^{'\star }_{ijJ'} \l '_{ijJ},  \  \l^{\star }_{ijJ'} \l _{ijJ},  \  
\l^{'\star }_{iJ k} \l '_{iJ'k}$, respectively, where  the
fermions generation index amongst the
dummy indices pairs, $(ij), \  (ik)$, refers to the third family.
Non vanishing contributions to ${\cal A}_{JJ'}$  could arise in Case {\bf C} if 
one assumed that two pairs of the above coupling constants products  
with different sfermions indices dominate, and   further requiring that these sfermions are
not mass degenerate. Another interesting possibility is
by assuming that  the hypothesis of  single pair of RPV coupling constants dominance 
applies  for the  fields current basis.
Applying then to the quark superfields
the transformation matrices relating these to mass basis fields, say, in the 
distinguished choice \cite{agashe} where the flavor changing effects bear on 
u-quarks, amounts to perform the substitution, $\l ' _{ijk} \to 
\l ^{'B} _{ink} V^\dagger _{nj}  $, where $ V$ is the CKM matrix.  The CP-odd 
factor, for the d-quark case, say, acquires then the form, 
$Im (l_{JJ'}^{ij \star } l_{JJ'}^{ij'})  \to 
\vert \l ^{'B} _{inJ'} \l ^{'B\star } _{imJ}\vert ^2  Im((V^\dagger )_{nj}
(V^\dagger )^\star _{mj} (V^\dagger )^\star _{nj'} (V^\dagger )_{mj'}),$
where the second factor on the right-hand side is recognized as the familiar
plaquette term, proportional to the products of sines of  all the 
CKM rotation angles times that of the CP-odd phase.

It may be useful to examine the bounds  on the RPV coupling constants 
implied by the current experimental limits on the 
flavor non diagonal leptonic widths \cite{branch2}, 
$B_{JJ'}^{exp} < [1.7,\ 9.8, \ 17. ] \ 10^{-6}$ for  the family 
couples, $[JJ'= 12,\ 23,\ 13]$.
The contributions associated with the $\l $ interactions can be  directly
deduced  from the results in Table \ref{table1}. Choosing the value, 
$\tilde m =100 $ GeV,  and  writing our numerical result
as, $ B_{JJ'}\approx ( {\l_{ijJ} \l _{ijJ'}^\star  \over 0.01 })^2  4 \ 10^{-9}$, then
under the hypothesis of a pair of dominant coupling constants, 
one deduces, $\l_{ijJ}\l_{ijJ'}^\star < [0.46,\ 1.1, \ 1.4 ]$, for  all fixed  choices 
of the family couples, $i,  \ j$.  (An extra factor $2$ in $B_{JJ'}$ 
has been included to account for the antisymmetry property of $\l_{ijk}$.)
For the $\l '$ interactions, stronger bounds obtain because of the 
extra color factor and of the internal top-quark  contributions. 
A numerical  calculation (not reported in Table \ref{table1}) 
performed with the choice, $\tilde m =100 GeV $ for Case {\bf C}, gives us:
$B_{JJ'}\approx ({ \l ^{'\star } _{Jjk} \l'_{J'jk} \over 0.01 })^2 1.17 \ 10^{-7}$,
which,  by comparison with the experimental limits, yields the bounds: 
$\l ^{'\star } _{Jjk} \l'_{J'jk} < [0.38,\ 0.91, \ 1.2 ] 10^{-1}$, for the same 
family configurations,  $[J \ J'= 12,\  23,\  13] $, as above. 
These  results agree  in size to within a factor of 2 with results
reported in a recently published  work \cite{anwar}.
%\vfill\eject
\begin{table}
\caption
{Flavor changing
rates and CP asymmetries for  d-quarks, leptons  and u-quarks pair production in 
the three cases,  appearing in line entries as Cases   {\bf A}, {\bf B} 
and {\bf C}, which correspond to internal lines belonging to all three families, the second and third families 
and the third family, respectively.     
The results for d-quarks and leptons, unlike those for up-quarks, 
are obtained in the approximation where one
neglects the final fermions masses.
The first four column fields (Z-pole column entry)  show results for
the Z-boson pole  branching fractions $B_{JJ'} $  and 
asymmetries, ${\cal A}_{JJ'}$. The last four column fields  (off Z-pole column entry) 
show results for the flavor non-diagonal cross sections,  $\sigma_{JJ'}$,
in fbarns  and for  the asymmetries, ${\cal A}_{JJ'}$, with photon and Z-boson
exchanges added in. The results in the  two lines for the off Z-boson pole 
are associated to the two values for the center of mass energy,
$s^{1/2} = 200, \ 500$ GeV. The columns subentries indicated by $\tilde m$
correspond to the sfermions mass parameter, $\tilde m=100, 
\ 1000 $  GeV. The notation $ d - n$ stands for $ \ 10^{-n}$. }
\vskip 0.2cm
\begin{tabular}{|c|cccc|cccc|}
\hline 
 & Z-pole & &&&  Off Z-pole & & &  \\
\cline{2-9}\cline{2-9} 
 & $\tilde m =100 $ & &  $\tilde m =1000$ & 
 & $\tilde m =100 $ & &  $\tilde m =1000$ &   \\
\cline{2-9} 
 & $ B_{JJ'}$& ${\cal A}_{JJ'} $& 
$ B_{JJ'}$& ${\cal A}_{JJ'} $& $\sigma_{JJ'} $ & ${\cal A}_{JJ'} $ &
$\sigma_{JJ'} $ & ${\cal A}_{JJ'} $ \\
\hline
${\bf d_J \bar d_{J'} }$ \\
    {\bf  A} & $ 5.6d-9$& $ 0.38$& $4.2d-11$& $0.068$& $11.5$& $-5.50d-3$& 
$1.46d-2$& $-5.72d-3$  \\
    & & & & & $3.62$& $-2.90d-3$& $6.90d-2$& $+3.17d-3$  \\
   \\
    {\bf  B} & $ 4.68d-9$& $ 0.20$& $4.12d-11$& $0.034$& $11.5$& $-6.80d-3$& 
$1.46d-2$& $-7.14d-3$  \\
    & & & & & $3.62$& $-3.47d-4$& $6.90d-2$& $+1.32d-3$  \\
   \\
    {\bf  C} & $ 3.8d-9$& $ 0.0$& $3.98d-11$& $0.0$& $11.5$& $-8.11d-3$& 
$1.46d-2$& $-8.55d-3$  \\
    & & & & & $3.62$& $+2.20d-3$& $6.90d-2$& $-5.18d-4$  \\
\hline 
    ${\bf l_J^- l^+_{J'} }$ \\
    {\bf  A} & $ 3.2d-9$& $ -0.44 d-3$& $3.6d-12$& $0.049$& $38.2$& $-1.18d-3$& 
$3.84d-2$& $-1.61d-3$  \\
    & & & & & $4.57$& $-6.90d-3$& $3.67d-1$& $-1.04d-3$  \\
\\
    {\bf  B} & $ 1.3d-9$& $ -0.55d-3$& $1.5d-12$& $-0.54d-3$& $38.2$& $-7.90d-4$& 
$3.84d-2$& $-1.08d-3$  \\
    & & & & & $4.57$& $-4.60d-3$& $3.67d-1$& $-6.93d-3$  \\
\\
   {\bf C} & $ 6.53d-10$& $ 0.0$& $7.5d-13$& $0.0$& $38.2$& $-3.95d-4$& 
$3.84d-2$& $-5.38d-4$  \\
   & & & & & $4.57$& $-2.30d-3$& $3.67d-1$& $-3.46d-4$  \\
\hline 
    ${\bf u\bar c }$ \\
    {\bf  A} & $ 6.5d-9$& $-0.69d-3 $& $ 8.9d-12$& $-0.12$& $ 11.5$& $2.63d-3 $& 
$1.46d-2$& $2.96d-3$  \\
    & & & & &$ 3.62$& $1.04d-2$& $6.90d-2$& $6.62d-3$  \\
\\
    {\bf  B} & $ 2.56d-9$& $- 0.89 d-3 $& $ 3.8d-12$& $-0.11$& $ 11.5$& $1.76d-3 $& 
$1.46d-2$& $1.98d-3$  \\
    & & & & &$ 3.62$& $6.93d-3$& $6.90d-2$& $4.42d-3$  \\
\\
    {\bf  C} & $ 1.26d-9$& $0.0 $& $ 1.95d-12$& $0.0$& $ 11.5$& $8.90d-4 $& 
$1.46d-2$& $1.00d-3$  \\
& & & & &$ 3.62$ & $3.47d-3$& $6.90d-2$& $2.21d-3$  \\
\hline 
${\bf t\bar c }$ \\
  {\bf  A} & $ $& $ $& $ $& $ $& $ 5.66$& $2.15d-3 $&$1.60d-3$& $3.31d-3$  \\
& & & & &$ 4.29$& $7.02d-3$& $5.95d-2$& $ 6.56d-3 $  \\
\\
   {\bf  B} & $ $& $ $& $ $& $ $& $ 5.66$& $1.43d-3 $&$1.60d-3$& $2.22d-3$  \\
& & & & &$ 4.29$& $4.68d-3$& $5.95d-2$& $ 4.38d-3 $  \\
\\
   {\bf  C} & $ $& $ $& $ $& $ $& $ 5.66$& $7.22d-4 $&$1.60d-3$& $1.13d-3$  \\
& & & & &$ 4.29$& $2.34d-3$& $5.95d-2$& $ 2.19d-3 $  \\
\\
\hline
\end{tabular}
\label{table1}
\end{table}
\subsubsection{Fermion anti-fermion pair production rates}
Let us now proceed to the off Z-boson pole observables.
The numerical results for
the flavor non-diagonal integrated cross sections and CP asymmetries 
are shown in Table \ref{table1} for two selected values
of the center of mass energy, $\sqrt s =  200 $ and $  500 $ 
GeV.  The numerical  results displaying  the 
variation of these observables with the center of  mass of energy  (fixed $\tilde m$) and with the superpartners mass parameter (fixed $\sqrt s$) 
are given in  Fig.\ref{fig4} and   Fig.\ref{fig5p}, respectively.
All the results presented in this work 
include both photon and Z-boson exchange contributions. 
We observe  here that the predictions for  asymmetries are 
sensitive to the interference effects between photon and Z-boson
exchange contributions.

We discuss first the predictions for flavor  non diagonal rates. We 
observe a strong decrease with increasing values of $\tilde m$ and a 
slow decrease with increasing values of $\sqrt s$. Following a rapid initial 
rise at threshold, the rates settle at values ranging between
$  (10 \ - 10^{-1})$  fbarns for a wide interval of $\tilde m$ values.  
The  dependence on $\tilde m$ can be approximately represented as,
$ \s_{JJ'}/ [  \vert {t^i_{JJ'}\over 0.01} \vert ^2 ({100 \over \tilde m})^{2 \ - \ 3} ] 
\approx  (1 \ - \ 10) $ fbarns  $ \approx R 
({\sqrt s  \over ( 1 \ TeV })^2 (10^{-1} \ - \ 1)  .$ 
The rate of decrease of $\s _{JJ'}$ with $\tilde m$ 
slows down with increasing $s$. It is   interesting to note
that if we had considered here  constant values of
the product $\l  \ (\tilde m/100 GeV) $, 
 rather than constant values of  $\l $, the 
 power dependence of rates  on  $\tilde m$ would  be such as to lead to
interestingly  enhanced rates at large $\tilde m$.

The marked differences exhibited by the results for lepton pair production, apparent on windows  $(c)$  and $(d)$ in Figures \ref{fig4} and \ref{fig5p} 
are due to our deliberate choice of adding  
the s-channel $\tilde \nu $ pole term   for the lepton case  while omitting it 
for the d-quark case.   The larger rates found  for leptons as compared 
to d-quarks, in spite of the extra color factor present  for d-quarks 
(recall that the $l^+l^-\to 
f_J\bar f_{J'} $ reactions rates  for down-quarks  and up-quarks 
pick up an extra color factor $ N_c$ with respect to those for  leptons) 
is  thus  explained by the strong enhancement induced by adding in
the sneutrino exchange contribution.
This choice  was made here 
for illustrative purposes, setting for orientation the relevant coupling constant  at the value, $\l_{1JJ'}=0.1$.
The  $\tilde \nu $ propagator pole  was 
smoothed out by employing  the familiar  shifted propagator 
prescription,  $( s-m^2_{\tilde 
\nu } +im_{\tilde \nu } \G_{\tilde \nu })^{-1}$, while describing approximately
 the   sneutrinos decay width in terms of the RPV contributions alone,
namely,  $\G (\tilde \nu_i \to l^-_k+ l^+_j)= { \l_{ijk}^2 {\tilde  m}_i
\over 16\pi }$  and 
$\G (\tilde \nu_i \to d_k+ \bar d_j)=N_c { {\l  '}_{ijk}^2 {\tilde  m}_i
\over 16\pi }$. 
%%%%%%%%%%%%%%%%%%%%%%%%%%%%%%%%%%%%%%%%%%%%%%%%
\begin{figure} [h]
\begin{center}
\leavevmode
\psfig{figure=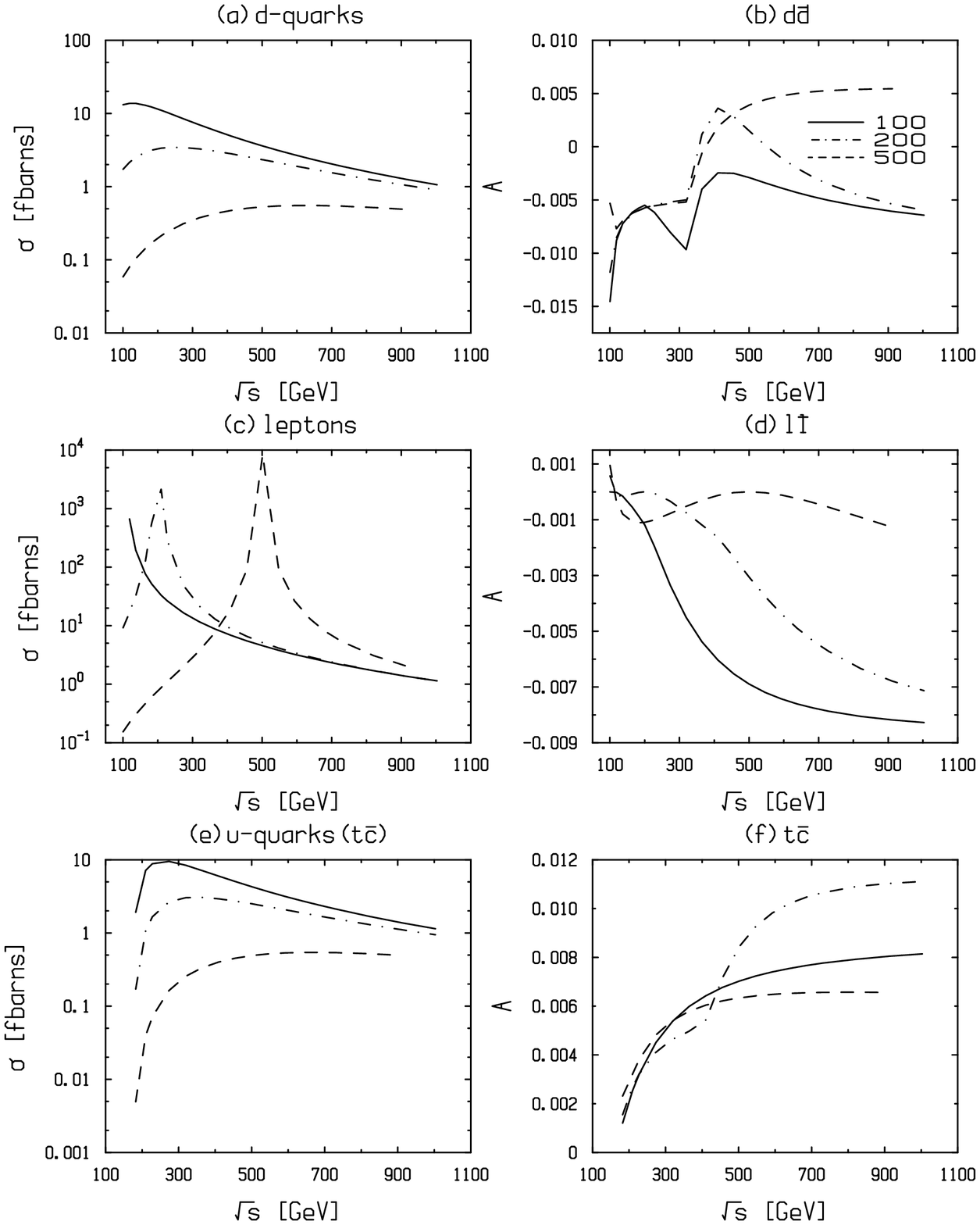}
\end{center}
\end{figure} 
\begin{figure}
\caption{Integrated flavor non-diagonal cross sections  
(left hand side windows) and asymmetries  (right hand side windows) as functions of the 
center of mass  energy in the production of 
down-quark-antiquark pairs (two upper figures  $(a)$ and $(b)$), 
lepton-antilepton pairs (two intermediate figures  $(c) $and $(d)$)
and up-quark-antiquark pairs of type $\bar t c+\bar c t$ 
(two lower figures (e) and (f)). 
The tree level amplitude 
includes only the t-channel contribution for the d-quark case, 
both t- and s-channel exchange contributions for the lepton 
case, and the u-channel exchange for the up-quark case. The one-loop 
amplitudes, with both  Z-boson  and photon exchange contributions, 
include all three internal fermions 
generations, corresponding  to Case {\bf A}. 
Three choices for the 
superpartners uniform mass parameter,  $\tilde m$, 
are considered: $ 100 GeV$ (continuous lines),
$ 200 GeV$ (dashed-dotted lines),  
$ 500 GeV$ (dashed lines).   }
\label{fig4}
\end{figure}
%%%%%%%%%%%%%%%%%%%%%%%%%%%%%%%%%%%%%%%%%%%%%%%%%
\begin{figure} [h]
\begin{center}
\leavevmode
\psfig{figure=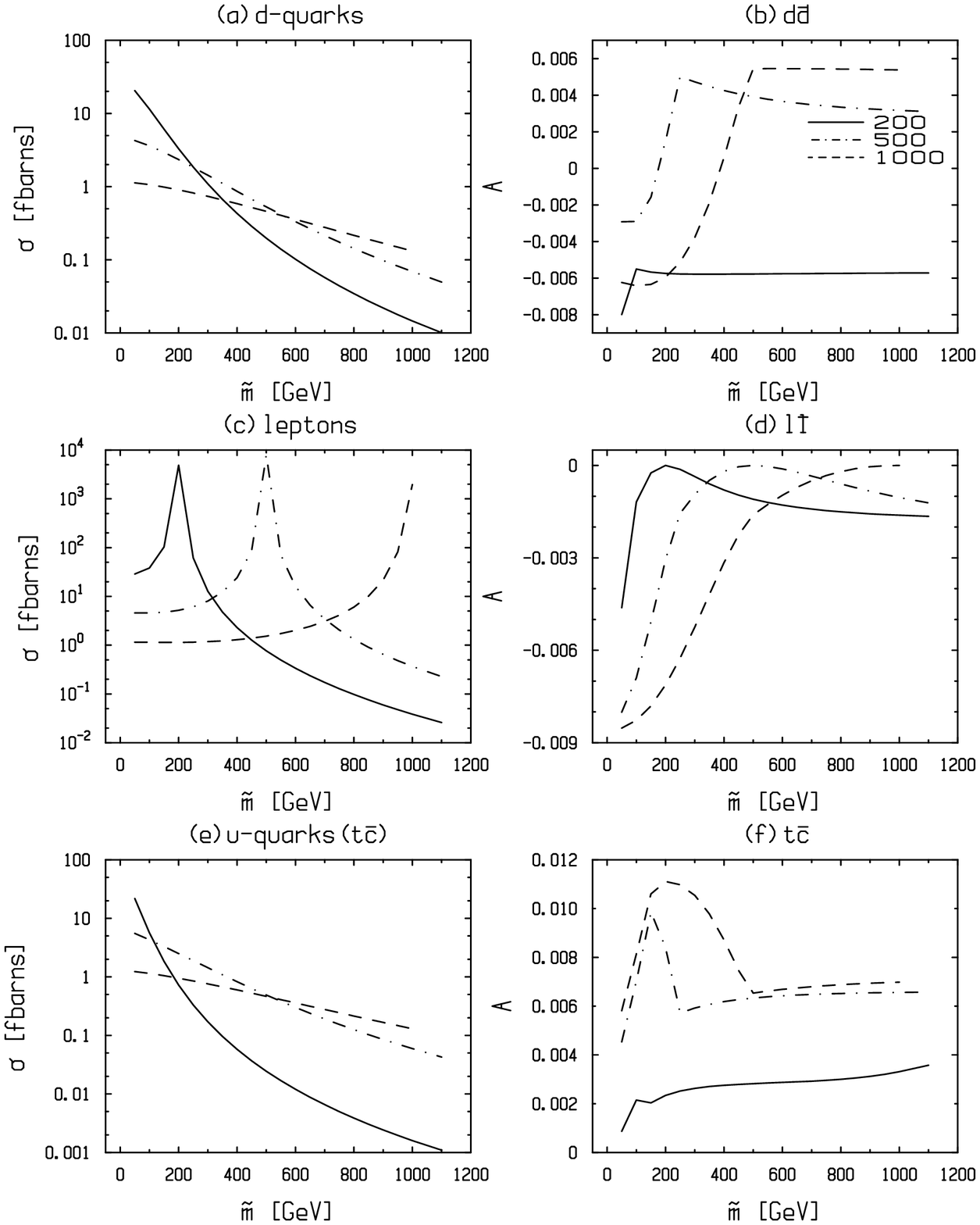}
\end{center}
%\vfill\eject
\caption{Integrated flavor non-diagonal cross sections  (left hand side windows)  and  CP-odd 
asymmetries (right hand side windows)  as functions of the 
scalar superpartners mass parameter, $\tilde m$, in the production of 
down-quark-antiquark pairs (two upper figures  $(a)$ and $(b)$), 
lepton-antilepton pairs (two intermediate figures  $(c) $and $(d)$)
and up-quark-antiquark pairs of type $\bar t c  $ or $ \bar c t$ 
(two lower figures (e) and (f)). 
The tree level amplitude 
includes only the t-channel contribution for the d-quark case, 
both t- and s-channel exchange contributions for the lepton 
case, and the u-channel exchange for the up-quark case. The one-loop 
amplitudes, with  both  photon and Z-boson exchange contributions,
include all three internal fermions 
generations, corresponding  to Case {\bf A}, with three families running inside loops. 
Three choices for the  center of mass energy, $s^{1/2}$, 
are considered: $ 200 GeV$ (continuous lines),
$ 500 GeV$ (dashed-dotted lines),  
$ 1000 GeV$ (dashed lines).   }
\label{fig5p}
\end{figure}
%%%%%%%%%%%%%%%%%%%%%%%%%%%%%%%%%%%%%%%%%%%%%%%%% 

Proceeding next to the CP-odd asymmetries, we note that since these
 scale as a function  of the RPV coupling constants  as, 
$ Im (l^{ij}_{JJ'} l^{i'j'\star }_{JJ'} )/\vert t^{i''}_{JJ'}
\vert ^2$, our present predictions are independent of 
the uniform reference  value assigned to these coupling constants. 
If the generational dependence of the 
RPV coupling constants  were to exhibit strong hierarchies, this peculiar 
rational dependence could induce strong suppression or enhancement factors.

The cusps in the dependence of ${\cal A}_{JJ'}$ on $\sqrt s$ (Fig. \ref{fig4}) 
occur at values of the center of mass energy where one crosses thresholds for 
fermion-antifermion (for the energies under consideration, $t\bar t$) pair production, 
$\sqrt s =2m_f$, and  scalar superpartners pair production, 
$\sqrt s =2\tilde m$. These are the thresholds for the processes,
$ l^-+l^+\to f\bar f $ or $l^-+l^+\to \tilde f' \tilde f^{'\star }$, 
at which the associated loop  amplitudes acquire finite imaginary parts. 
Correspondingly, in the dependence of  ${\cal A}_{JJ'}$ on $\tilde m$ 
(Fig. \ref{fig5p}) the cusps appear at $\tilde m =\sqrt s /2$. 
We note on the results  that the $t\bar t$  
contributions act to suppress the asymmetries whereas 
the $\tilde f\tilde f^\star $ contributions  rather act to  enhance them. 
Sufficiently beyond these two-particle thresholds, the asymmetries vary
weakly with $\tilde m$.
A  more rapid variation as a function of energy occurs
in the leptons production case
due to the addition there of the sneutrino pole contribution. 

The comparison  of results for asymmetries in  Cases {\bf A, \ 
B, \ C} reflects on the dependence of loop integrals  
with respect to the internal fermions masses.
An examination of Table \ref{table1} reveals that for leptons and
up-quarks, where intermediate states involve leptons or d-quarks,
all three families have nearly equal contributions. 
The results for down-quarks production  are
enhanced because of  the intermediate  top-quark  contribution, which dominates over 
that of lighter families. However, this effect is depleted when the  finite 
imaginary part from $t\bar t$ sets in.
The asymmetries for up-quarks production  
assume values in the range, $10^{-2}- 10^{-3}$,
irrespective of the fact that the final fermions belong to light 
or heavy families.   
\section{Conclusions}
\label{seconc}
The two-body  production at high energy leptonic colliders
of fermion pairs of different families 
could  provide valuable information on the flavor structure of the
R parity odd Yukawa interactions.  One can only wish that  an  experimental  
identification of    lepton and quark flavors  at high energies 
becomes accessible in the future.
Although the  supersymmetric  loop corrections to these processes 
may not be as strongly  
suppressed as their  standard model counterparts,  
one expects that the  degeneracy or alignment constraints on  the scalar 
superpartners masses and flavor mixing  should 
severely bound their contributions.
Systematic studies  of the supersymmetry corrections  to the 
 flavor changing rates and CP asymmetries in fermion pair production should be 
strongly encouraged.
 
An important characteristic of the R parity odd interactions  is that 
they can  contribute to integrated  rates 
at tree level and to CP asymmetries through 
interference terms between the  tree and loop amplitudes. While we have 
restricted ourselves to the subset of loop contributions 
associated with Z-boson exchange,  a large number of 
contributions,  involving quark-sleptons  or 
lepton-squarks intermediate states in various families configurations,
could still  occur. 
The contributions to rates and asymmetries depend
strongly  on the values of the 
R parity odd  coupling constants. Only the rates   are directly 
sensitive   to  the supersymmetry  breaking scale.
To  circumvent the uncertainties from the sparticles spectrum, we have
resorted to the simplifying   assumption that  the scalar 
superpartners mass differences and mixings 
can be neglected. We have  set the  RPV coupling constants  at
a uniform value while sampling a set of  cases
from which one might reconstruct the family dependence of the RPV coupling constants. We have also embedded  a CP complex
phase  in the RPV coupling constants in a specific way, meant to serve
mainly as an illustrative 
example. Although  the  representative  cases that we have considered 
 represent a small fraction of  the host of  possible variations, 
they give a fair idea of the sizes to expect.
Since these processes cover a wide range of family configurations, one 
optimitistic  possibility  could be that  one specific entry for 
the family configurations  would enter with 
a sizeable RPV coupling constant. 

The contributions  to the  flavor changing rates have a 
strong sensitivity on the  RPV coupling constants and the superpartners  mass,
 involving high powers of these  parameters.
We find a generic dependence for the 
flavor changing  Z-boson decay branching ratios  of form, 
$({\l \l  \over 0.01})^2 ({100\over \tilde m})^{2.5} \ 10^{-9}$.
For the typical bounds on the RPV coupling constants,
it  appears that these branchings  are
three order of magnitudes below the current experimental sensitivity.  
At higher energies, the 
flavor changing rates are in  order of magnitude, 
$({\l \l  \over 0.01})^2 ({100\over \tilde m})^{2\ -\  3 }  \ (1\ - 10 )$ fbarns. 
Given  the size for the typical   
integrated luminosity, ${\cal L}= 50 fb^{-1}$/year, 
anticipated  at the future leptonic  machines, one can be moderately optimistic
on the observation of clear signals. 

The Z-boson pole  CP-odd 
asymmetries  are of order,  $ (10^{-1} \ -\ 10^{-3}) \sin \psi $.
For the off Z-boson pole reactions, a CP-odd  phase, $\psi $,  embedded in the 
RPV coupling constants shows up in  asymmetries with  reduced strength, 
$(10^{-2} - 10^{-3})\sin \psi $ for leptons,
d-quarks and u-quarks. 
The largely unknown structure of the RPV coupling constants in flavor space
leaves room for good or bad surprises,  since the 
peculiar rational dependence  on the coupling constants,
$ Im (\l \l ^\star  \l \l  ^\star )/\l ^4 $, 
and similarly with $\l \to \l '$,  may lead to strong enhancement or suppression factors. 

%\section{TABLES}
%\vfill\eject
%\newpage
%\footnotesize
%\begin{references}
%{\bf References}

\end{document}